\documentclass[12pt]{article}
\usepackage{etex}
\usepackage{amsfonts}
\usepackage{latexsym}
\usepackage{natbib}
\usepackage{pst-pdf, pst-all, amsthm, amssymb, latexsym}
\usepackage{amsmath}
\usepackage{comment}
\usepackage{listings}
\usepackage{pstricks,pstricks-add} 
\usepackage{booktabs}
\usepackage{rotating}
\usepackage{multirow, tabularx}
\usepackage{graphicx}
\usepackage{lscape}
\usepackage{abstract}
\usepackage{lipsum} 
\usepackage{ctable}
\def\sym#1{\ifmmode^{#1}\else\(^{#1}\)\fi}
\usepackage{subfloat}
\usepackage{titlesec}
\usepackage{blindtext}
\usepackage{rotating}

\usepackage[margin=2cm]{caption}
\usepackage[margin=3.6cm]{subcaption}
\usepackage{mathtools}

\AtBeginDocument{%
  \addtolength\abovedisplayskip{-0.15\baselineskip}%
  \addtolength\belowdisplayskip{-0.15\baselineskip}%
  }
  \usepackage{lmodern}

\usepackage{fnpct}

\makeatother
\usepackage{indentfirst}

\usepackage{enumerate}
\renewcommand{\labelitemi}{$\square$}
\usepackage{hyperref}
\hypersetup{backref,  
pdfpagemode=FullScreen,  
colorlinks=true, citecolor=blue!50!black}
\usepackage{graphicx}

\usepackage{titlesec}

\titleformat{\paragraph}
{\normalfont\normalsize\bfseries}{\theparagraph}{1em}{}
\titlespacing*{\paragraph}
{0pt}{3.25ex plus 1ex minus .2ex}{1.5ex plus .2ex}

\usepackage{titlesec}
\titlespacing*{\section}{0pt}{0.9\baselineskip}{0.7\baselineskip}
\titlespacing*{\subsection}{0pt}{0.7\baselineskip}{0.7\baselineskip}
\titlespacing*{\subsubsection}{0pt}{0.7\baselineskip}{0.7\baselineskip}

\usepackage{fancyhdr}

\newtheorem*{prop1}{Proposition 1} 
\newtheorem*{prop1a}{Proposition 1a} 
\newtheorem*{prop2}{Proposition 2} 
\newtheorem*{prop2a}{Proposition 2a} 
\newtheorem*{prop3}{Proposition 3} 
\newtheorem*{prop3a}{Proposition 3a} 
\newtheorem*{prop4}{Proposition 4} 
 
\newtheorem*{prop5}{Proposition 5}

\usepackage{scalerel}
\newcommand\vfrac[2]{\ThisStyle{%
  \setbox0=\hbox{$\SavedStyle#1#2$}%
  \setbox2=\hbox{$\SavedStyle X$}%
  \ifdim\ht0>\ht2\setlength{\ht0}{\ht2}\fi%
  #1\mathord{\stretchto{\raisebox{2.3\LMpt}{$\SavedStyle/$}}{\ht0}}#2}}

\titleformat{\subsubsection}
  {\normalfont\fontsize{12}{17}\selectfont}{\thesubsubsection}{1em}{}
 
\usepackage{enumitem}

\usepackage{cleveref}
\usepackage{subcaption}

\usepackage{relsize,etoolbox}
\AtBeginEnvironment{quote}{\smaller}

\begin{document}

\title{External Threats, Political Turnover and \\ Fiscal Capacity}
\author{Hector Galindo-Silva\thanks{Department of Economics, Pontificia Universidad Javeriana. Email: galindoh@javeriana.edu.co. I thank Guillermo Diaz, Ruben Enikolopov, Raphael Godefroy, David Karp, Nicolas Lillo, Didac Queralt and Alessandro Riboni for their extremely helpful comments and suggestions. I am also grateful for the hospitality of the Barcelona Institute for Political Economy and Governance (IPEG) at the Universitat Pompeu Fabra, \'Ecole Polytechnique and the University of Montreal, where part of this work was written. Any remaining errors are my own.}\\
Pontificia Universidad Javeriana
}

\date{This version: January 2020}
\maketitle

\vspace{-0.5cm}

\begin{abstract}

In most of the recent literature on state capacity, the significance of  wars in state-building assumes that threats from foreign countries generate common interests among domestic groups, leading to larger investments in state capacity. However, many countries that have suffered external conflicts don't experience increased unity. Instead, they face factional politics that often lead to destructive civil wars. This paper develops a theory of the impact of interstate conflicts on fiscal capacity in which fighting an external threat is not always a common-interest public good, and in which interstate conflicts can lead to civil wars. The theory identifies conditions under which an increased risk of external conflict decreases the chance of civil war, which in turn results in a government with a longer political life and with more incentives to invest in fiscal capacity. These conditions depend on the cohesiveness of institutions, but in a non-trivial and novel way: a higher risk of an external conflict that results in lower political turnover, but that also makes a foreign invasion more likely, contributes to state-building only if institutions are sufficiently incohesive.


\end{abstract}

\newpage


\section{Introduction}

There is a large and growing literature on the impact of war on state-building.\footnote{\label{litstatecap}See \cite{Tilly1975, Tilly1990} and, more recently,  \cite{BesleyPersson2008JEEA, BesleyPersson2009AER, BesleyPersson2010ECTA, BesleyPersson2011}. Also see \cite{Dincecco2011}, \cite{DinceccoPrado2012}, \cite{OBrienCasalilla2012}, \cite{GennaioliVoth2015} and  \cite{KoKoyamaSng2018} for European cases, and \cite{LopezAlves2000}, \cite{Centeno2002} and \cite{Thies2005} for Latin America.} Much of this literature builds on Charles Tilly's famous phrase, ``War made the state and the state made war'' \citep[p. 42]{Tilly1975}.
In most of this literature, which includes an important series of papers that attempt to unify some essential theories about state-building \citep[which I will refer to from now on as \emph{B\&P}]{BesleyPersson2008JEEA, BesleyPersson2009AER, BesleyPersson2010ECTA}, the significance of war relies on the assumption that threats from foreign countries generate common interests among domestic groups, leading to larger investments in state capacity.

This paper develops an alternative theory of the impact of interstate conflicts on fiscal capacity, in which fighting an external threat is not always a common-interest public good, and in which interstate conflicts can lead to civil wars. The theory identifies conditions under which an increased risk of external conflict decreases the chance of civil war, which in turn results in a government with a longer political life and with more incentives to invest in fiscal capacity. 
 
 The idea that interstate wars have a positive effect on state-building because of their contribution to the provision of a public good (e.g. national defense) has helped explain many crucial fiscal innovations in Europe from the 17\textsuperscript{th} to 19\textsuperscript{th} centuries.\footnote{See \cite{HoffmanRosenthal1997}, \cite{Dincecco2011} and \cite{OBrienCasalilla2012}.} However, important issues remain. For instance, \cite{GennaioliVoth2015} show that during the period of initial European state building, interstate warfare was mostly a private good for princes in pursuit of glory and personal power. \cite{PincusRobinson2014} argue that this thesis does not apply to Britain, noting that critical elements of state-building (such as a monopoly on violence) were not associated with interstate wars but rather were either uncorrelated with wars or associated with civil wars.
 
\begin{figure}[h!]
\begin{center}
\caption{Civil wars and interstate conflicts (partial correlation)}\label{civwardisputesall}
\resizebox{11cm}{7cm}{\includegraphics[width=4in]{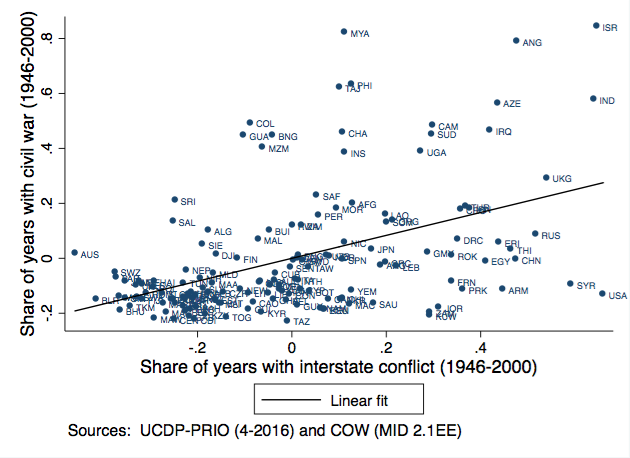}}
\end{center}
\end{figure}

 If we extend the hypothesis to other regions and more recent times, the idea that interstate conflicts generate common interests among groups seems even less plausible. In the last century, many countries that experienced external conflicts were also affected by factional politics that drove them to destructive civil wars. Figure \ref{civwardisputesall} shows the partial correlation between civil wars and interstate conflicts by plotting the share of years with a civil war against the share of years with an interstate dispute between 1946 and 2000.\footnote{The data on civil wars is from the UCDP/PRIO. The data on interstate disputes is from the Correlates of War (COW) project, and measures whether a given country is engaged in a militarized interstate dispute (MID) of high intensity (with at least a display of force) in a given year. The underlying regression controls for executive constraints (between 1946 and 2000), ethnic fractionalization and legal origin.} The figure shows a significant positive correlation, meaning that countries that experienced more interstate disputes also experienced more civil wars. This pattern is confirmed when we look at each interstate dispute and civil war in detail. Between 1946 and 2000, 62\% of countries that experienced an interstate dispute also experienced a civil war. In addition, more than 67\% of civil wars occurred within a window of plus/minus two years surrounding an interstate conflict.\footnote{This percentage is lower if instead of looking at high intensity militarized interstate disputes, we look at interstate ``wars,'' defined as militarized interstate disputes with a minimum of 1,000 battle-related combatant fatalities within a 12-month period (see the COW project). This appears to be the definition used by \emph{B\&P} in their empirical analysis. However, even when we focus on these very high-intensity conflicts, at least 12\% of civil wars between 1946 and 2000 occurred  within a window of plus/minus two years surrounding an interstate war.} 
 
An abundant literature on conflict has documented close links between civil wars and conflicts between states that suggests there might be an important causal relationship.\footnote{See \cite{GleditschBeardsley2004, HegreSambanis2006, Gleditsch2007, Salehyan2008, GleditschSalehyanSchultz2008, Cunningham2010, GleditschSalehyanSkredeCunningham2011}; and \cite{MorelliPischedda2013}.} These examples include post-independence and Cold War periods in Latin America, sub-Saharan Africa and southeast Asia. Two notable examples, which I discuss in more detail later, are Mexico following its independence, and the Democratic Republic of the Congo (DRC) during Mobutu Sese Seko's rule. In both cases, the risk of an external conflict played a crucial role in the occurrence, intensity and persistence of internal conflicts that these countries experienced during several stages of their history.  Other examples include the risk of a US invasion seeking to prevent  ``a second Cuba'' in Chile , which played a role in the 1973 Chilean coup d'\'etat \citep[see][chs. 11-13]{CollierSater2004}, and the risk of a US trying to overthrow the ``Marxist Sandinistas'' in Nicaragua, which affected the Nicaraguan Contra conflict \citep[see][]{PrevostVanden1999}.
 
If the risk of an external conflict can potentially affect whether a country suffers a civil war, the hypothesis that interstate conflicts generate common interests among domestic groups, resulting in more state-building, should be re-examined.  This paper provides an alternative explanation of the impact of interstate conflicts on state capacity that accounts for the possibility that the likelihood of a civil war depends on the risk of an interstate conflict, and that does not assume that fighting an interstate conflict is always a common-interest public good.\footnote{Tilly considers the possibility of a close relationship between interstate disputes and civil wars in scenarios in which the nationalization and specialization of the military was very weak \cite[see][p. 186]{Tilly1990}. He associates this possibility with European countries before dynasties typically controlled states, and does not directly relate this idea with the analysis of the impact of war on state-building.} My aim is to shed light on the process of state-building (or lack thereof) when countries experience a high risk of an external conflict, as well as the potential for a related civil war. 

The paper focuses on fiscal capacity, understood as the capacity of a state to generate tax revenue.  It uses  \emph{B\&P}'s basic framework, which is extended and modified in several respects. It starts by proposing a baseline scenario where a government can lose power for three reasons: by losing to a foreign country in an external conflict, or by losing to the domestic opposition in an internal conflict or an election.  Crucially, the opposition and government face different payoffs if the country loses in an external conflict. The opposition must decide whether to initiate a civil war, knowing that a civil war may also weaken the government against the foreign country. The government must decide whether or not to invest in fiscal capacity without knowing whether the opposition will trigger a civil war. The paper focuses on how an increased risk of external conflict affects the incentives of the government to invest in fiscal capacity. 

The main result  identifies conditions under which an increased risk of external conflict may either increase or decrease  investment in fiscal capacity. The outcome depends on whether the higher risk of external conflict provides the domestic opposition with enough incentives to support the government against the external threat.  If the heightened risk of an external conflict significantly strengthens the government domestically, so that it  outweighs the higher risk of a foreign administration, and the probability of political turnover decreases, it provides the government with incentives to invest in fiscal capacity insofar as a sufficiently low level of institutional cohesiveness allows it to extract rents. 

The conditions under which an increased risk of external conflict either increases or decreases  investments in fiscal capacity  implies  a new and interesting result: if as a result of an increased risk of external conflict, the chance that the incumbent stays in power increases, the government does not always decide to invest in fiscal capacity. The exact relationship depends on how cohesive institutions are.  This result contributes to the important debate about whether more political stability leads to better public policies \cite[see][]{AcemogluGolosovTsyvinski2011}, and provides a novel mechanism through which this may not be the case: when the greater political stability results from a higher risk of  an external conflict. 

The baseline model is then extended to include an endogenously chosen level of institutional cohesiveness.  Besides establishing the conditions under which there is a positive level of institutional cohesiveness and providing an explicit formula for this level, this extension confirms the results previously mentioned. 

The paper also examines two episodes from post-independence Mexico.\footnote{In Appendix A.1, I also discuss one episode  from the Democratic Republic of Congo (DRC) during Mobutu Sese Seko's rule.} These episodes illustrate some of the trade-offs that are formalized in the model. In particular, during certain critical periods of Mexico's history, government decisions about institutions were significantly affected by the possibility of a civil war, which in turn depended on the likelihood of a foreign intervention.

This study contributes to the literature in several ways. First, it expands on \emph{B\&P}'s framework in a novel way. It is similar to \cite{BesleyPersson2008JEEA} in that it also studies the impact of internal and external conflicts on fiscal capacity. They argue that investment in fiscal capacity is lower when there is a greater risk of a future \emph{internal} conflict, and higher when there is a greater risk of a future \emph{external} conflict. As external conflicts can also occur in the present, which reduces investment in state capacity, they argue that the possibility of external conflict has an ambiguous effect. However, and importantly, their model and results depend on internal and external conflicts being independent. In particular, they assume that internal conflicts only occur when there is no external conflict. My model removes this assumption. 
 
This paper also shares similarities with \cite{GennaioliVoth2015} and \cite{KoKoyamaSng2018}, which also study the impact of war on fiscal capacity. \cite{GennaioliVoth2015} focus on initial European state-building (1600-1800). By proposing a model and providing empirical evidence,  \citeauthor{GennaioliVoth2015} argue that war's impact on state capacity (which they define as a centralized revenue-collection system) depends on the cost of war and on the initial level of political fragmentation. In their model, war does not necessarily lead to state-building when the costs of war are sufficiently low and fragmentation is sufficiently high. In such a case, it is better for rulers not to invest in centralized revenue collection because it is expensive, they would have more to lose (in foregone tax revenue) in the event of defeat, and war can be a cheaper alternative. 

\cite{KoKoyamaSng2018}  focus on comparing long-term institutional development in China and Europe.  They develop a Hotelling-style model to show how one-sided and two-sided external threats affect rebellion, political fragmentation, military investments, fiscal viability and taxation. They find that a bigger external threat always leads to military investments and decreases the probability of rebellion; whether the threat is one- or two-sided affects the levels of political fragmentation, fiscal viability and taxation.

Although the design of \cite{GennaioliVoth2015} and \cite{KoKoyamaSng2018} share some similarities with my model, there is one crucial difference. In their models, groups that form a country or continent act as a single entity when facing a external threat, and fighting this threat is still a common-interest public good in the sense that losses are equally distributed when a country loses a war (\citeauthor{GennaioliVoth2015}) or a continent leaves a significant part of its territory unprotected (\citeauthor{KoKoyamaSng2018}). In my model, a country's government and opposition can, \emph{a priori}, support or oppose a foreign threat, which reinforces or diminishes the impact of the external threat. 

The structure of this paper is as follows. In the next section, I discuss the case studies, with a particular emphasis on how the risk of external conflicts affected decisions about institution-building. Section \ref{model} presents the model, and Section \ref{conclusion} concludes.


\section{Illustrative Cases}
\label{cases}

This section briefly discusses two episodes from post-independence Mexico, which illustrate how the risk of external conflict, and government and opposition responses to that risk, helps explain important aspects of Mexico's early state-building process. This section's objective is to set the stage for a possible mechanism, which will be formally developed in Section \ref{model}.\footnote{As previously mentioned, Appendix A.1 also discusses a Cold War-era case study from the Democratic Republic of Congo (DRC), which illustrates how the risk of external conflict helps to explain DRC's  state-building process (particularly the failure of that process). The fact that the formalization in Section \ref{model}  also helps to explain this episode, which comes from the post-war era, shows the generality of the model proposed in this paper: it could help explain both pre-war and post-war  conflicts, which, although they differ in important aspects (such as the level of control sought by the invading countries), also share crucial similarities. I thank a referee for this observation.} 

\smallskip

During the first  decades of its independence, Mexico experienced a number of episodes that exemplify two  previously described roles of external threats in the state-building process: a high risk of external conflict (with Spain in 1829 and the United States from 1846-48) followed by either a decrease in political turnover and the implementation of key fiscal reforms, or by an increase in political turnover and the lack of much-needed reforms. I focus on two episodes that exemplify these two possibilities: Vicente Guerrero's presidency in 1829 and  the Mexican-American war between 1846 and 1848.\footnote{For other periods and alternative mechanisms that emphasize the role of domestic factors in building fiscal capacity in Mexico, see \cite{Garfias2018APSR}  for post-revolutionary Mexico and \cite{Garfias2018JoP} and \cite{Arias2013} for late colonial Mexico.}


\subsection{Vicente Guerrero's Presidency}

 In April 1829, at the beginning of his mandate, Mexican president Vicente Guerrero encountered serious challenges, including a series of domestic political conflicts and an empty treasury.  A crucial challenge during his presidency was the Spanish government's second reconquest attempt in July 1829. The specific timing was motivated by an international environment favorable to the reconquest and rich emigrants' promises to pay for the invasion plans  \citep[p. 59]{Sims1984}.

Another crucial characteristic of Guerrero's presidency was the occurrence, in August 1829, of a temporary respite in domestic political conflicts because of the Spanish threat. As Jan Bazant explains, ``the long-awaited invasion by Spanish troops came at the end of July 1829, and it served to cause a temporary lull in the factional political conflict as the nation rallied to the call for unity" \citep[p. 433]{Bazant1985}. This respite was preceded by two mass  expulsions  of Spaniards, who were  believed to  ``represent a threat to Mexican independence"  \citep[p. 9]{Sims1990}.

In the context of these events, at the end of August 1829, Vicente Guerrero requested and obtained extraordinary powers from the Mexican Congress to enact a tax reform \citep[see][]{Serrano2002, SerranoVazquez2010}. As a result, a national income tax was established in September 1829, for the first time in the history of independent Mexico \citep{Serrano2005}. This fiscal reform constituted ``one of the most radical transformations to the tax structure inherited from colonial times" \citep[p. 273]{Serrano2005}.\footnote{See also \cite{Jauregui2003b, Jauregui2003, Jauregui2005} and \cite{Serrano2002}.} As Jose Serrano says: 

\begin{quote}
On August 18, President Guerrero informed Congress that he considered the constitutional powers to which the government could appeal to confront the Spanish invasion insufficient ... The Congress ... authorized the Executive Branch to adopt as many measures as are necessary  ... Under the protection of extraordinary powers, Zavala [Guerrero's minister of finance] promoted a broad reform of the country's tax system ... [under this law] lawyers, doctors, surgeons, scribes and all ``professionals'' would spend \$24, and civil and military employees, both from the Federation and from the states, would contribute a part of their salary \citep[p. 106]{Serrano2002}.\footnote{See also \citet[p. 417]{SerranoVazquez2010}.}
\end{quote}

\medskip

Although the income tax was reversed two years later, this case illustrates how the risk of an external conflict can result in investments in fiscal capacity: support from the domestic opposition, which in this case occurred through a respite in internal conflicts and was preceded by a mass expulsion of Spaniards, gave the government incentives to propose and implement key fiscal reforms. The model in Section \ref{model} will propose one way to formalize this mechanism that emphasizes the role of the decrease in political turnover in providing the government  with incentives to increase fiscal capacity.

\subsection{The Mexican-American war}

The Mexican-American war occurred from 1846 to 1848. It had its origins in the decree of the US Congress annexing Texas in February 1845, which also made explicit America's intention to control California and all territory north of the Rio Grande. A puzzling characteristic of this period was that, unlike during the second Spanish reconquest attempt, Mexicans were divided and lacked commitment to their country (\citealp[p. 40]{Vazquez1997}; \citealp[pp. 184-201]{Bringas2008}; \citealp[pp. 142-143]{Cardenas2015}). Enrique Cardenas mentions the relevance of the rebellion led by the commander of the reserve army, General Paredes, who 
``marched towards Mexico City ... to take power, instead of going to the border to reinforce the defense" \citep[p. 142]{Cardenas2015}.  Peter Guardino also notes this division in the context of late February 1847's civil war, in which

\begin{quote}
 [Mexican] National Guard units of differing political sympathies confronted each other in Mexico City beginning in late February 1847. Many Mexicans understood this as the worst moment of the war, one in which Mexico's fractious politics undercut its defense just as the Americans threatened the center of the country. The conflict made it much easier for the Americans to launch the invasion of central Mexico that eventually defeated the country \citep[p. 175]{Guardino2017}.
 \end{quote}

Numerous diaries of American soldiers also reported a constant astonishment at the warm welcome they received from the majority of Mexicans, who expressed their desire that the occupying troops remain \citep[pp. 185 and 190]{Bringas2008}. The reason seems to be that ``local populations thought that they would get rid of the abuses of the local military and receive economic benefits from the sale of various products or from providing services" \citep[p. 190]{Bringas2008}, and that ``the loyalties of the inhabitants, especially those living on the northern border, had been conditioned by a series of local and regional alliances, economic ties, and political interests that increasingly linked them to the United States" \citep{Resendez1997}.  

A second characteristic of this period was that the Mexican government decided to raise revenues ``by mortgaging or selling property belonging to various Catholic Church organizations" \citep[p. 176]{Guardino2017}, instead of introducing new direct taxes (as Vicente Guerrero had done two decades prior). The government and Congress emphasized how important it was that ``the executive will be authorized only to take over the assets of the Catholic Church" \citep[p. 64]{Sordo1997},
and the main reason seems to be, as Mexican vice-president Gomez Farias explained to Mexico's president, Antonio Lopez de Santa Anna, ``the distrust [associated with] the misuse of funds and ruinous contracts" \citep[p. 64]{Sordo1997}.  

Facing a high risk of invasion by the US and an associated increase in the risk of an internal conflict, the Mexican government opted for a policy through which it could  only take over the assets of the Catholic Church, instead of widespread and potentially state-building fiscal reforms. This policy was seen by the government and Congress as the best alternative given the distrust of the government's ability to use funds responsibly. 

\medskip
This second case illustrates another mechanism through which the risk of external conflict can affect investments in fiscal capacity. If the domestic opposition, in response to an increased risk of external conflict, decides not to support the incumbent government and instead exploits the situation and starts an internal conflict, then the increased chances of political turnover, as well as concerns about increased opportunities for rent extraction by a future government, could disincentivize the current government from implementing capacity-building reforms.


\section{Model}
\label{model}
\subsection{Basic Model Setup}

There are two countries, $D$ (domestic) and $F$ (foreign), and two time periods, $s=1,2$. Country $D$ is composed of two groups, $A$ and $B$, each of which make up half of the population in every time period.\footnote{This is without loss of generality for the main results, and  simplifies the exposition (see footnote \ref{difpopshares}).} $F$ is homogeneous, and is ruled by the same group in both periods, also denoted by $F$.
The analysis focuses on $D$. However, $F$ is crucial to the analysis: through the threat of an intervention, $F$ affects decision-making in country $D$.

At the beginning of period 1, one of the domestic groups in $D$ ($A$ or $B$)  is chosen at random to be the government.  The government decides on a set of policies to be implemented during this period. This includes a uniform income tax rate, $t_1$, which is applied to individuals from groups $A$ and $B$, and a set of group-specific transfers, $\textbf{r}_1=\{r_1^A,r_1^B,r_1^F\}$, awarded to $A$, $B$ and $F$. The government in $D$ also determines, through investment, the period-2 stock of fiscal capacity, $\tau_2$.\footnote{As will be specified below, this is a model of pure redistribution since individuals do not derive utility from the consumption of public goods, and government revenue is only spent on transfers.}

Let $I_1\in\{A,B\}$ be the government in period 1, and $O_1\in\{A,B\}$ its domestic opposition.  After $I_1$ chooses the period-1  policies and investment, $O_1$ decides whether or not to contest $I_1$'s leadership by triggering a civil war.  If $O_1$ decides not to trigger a civil war, elections occur. Irrespective of $O_1$'s decision, an interstate conflict between $F$ and $D$ occurs with probability $\alpha$, which I assume to be exogenous.\footnote{\label{alphaex}By assuming that the probability of an interstate conflict is exogenous and that the occurrence of a civil war depends only on the opposition's choice, the model does not explicitly allow for scenarios in which a civil war occurs mainly because of the support provided to the opposition by a foreign country. I abstract from these situations, and simply model them as interstate conflicts (the case from the DRC discussed in Appendix A.1 could fall into this category). However, the theory proposed in this paper is consistent with scenarios in which a foreign country provides military support to the opposition (or to the incumbent government), but the theory does not model them explicitly insofar as it does not provide a micro-foundation for the foreign country's actions. Endogenizing  $\alpha$ is beyond the scope of this paper.} This interstate conflict can end in an invasion in which $F$ is the new incumbent, or in $O_1$ taking power in period 2 as a result of $F$'s direct intervention. Conditional on $F$ winning the interstate war, let $\lambda$ denote the probability that $O_1$ takes power thanks to $F$, and $1-\lambda$  the probability that $F$  establishes a foreign administration.\footnote{\label{discussionlambda}A larger $\lambda$ captures scenarios in which a direct foreign intervention primarily seeks a new  distribution of power between the domestic actors, rather than the establishment of a foreign administration. In this regard,  and importantly, $1-\lambda$ can also be interpreted as the extent to which a government led by $O_1$ but that is in power thanks to $F$'s direct intervention, is a `puppet government'. These scenarios seem to have been common in the post-World War II period (see \citealt{Owen2002}). Some potential examples include the Soviet Union's 1979 intervention in Afghanistan (see \citealt[][p. 351-352]{Westad2005}, and \citealt{Saikal2010}), and the US interventions in the Dominican Republic in 1965 \cite[see][p. 151]{Westad2005} and Grenada in 1983 \cite[see][]{Williams2007}. A more recent example may be the US invasion of Iraq in 2003 \cite[for instance, see][]{Wilson2009}. Softer interventions such as Russian interference in the 2016 US presidential election \cite[see][]{nytimes062019} could also be represented by $\lambda$. I thank a referee for suggesting this possibility and for having identified some of these cases.} 

The outcome of the interstate conflict, the domestic dispute, and/or elections determines the government in the second period, denoted by $I_2\in\{A,B,F\}$. The case $I_2=F$ occurs if $F$ wins the interstate conflict, and directly rules $D$ by establishing a foreign administration. In the second period, $I_2$ decides on a new set of policies.\footnote{Note that by assuming that the model ends in period 2, and that in this period there is no civil war or interstate conflict, the model does not  allow for scenarios in which a civil war is a consequence of an interstate conflict that ended, for instance, in a foreign administration. The study of these scenarios will require generalizing the model to a dynamic setting, which is left for future work.} 

Political turnover occurs when $O_1$ wins an election or a civil war, or when $F$ establishes a foreign administration. I assume that conditional on the occurrence of a civil war, the probability of any of these events is exogenous.\footnote{\label{paramex}Endogenizing these probabilities is left for future work.  Perhaps the stronger assumption is that the model does not allow for $O_1$ and $I_1$ to invest resources to increase their chances of victory. Thus, taxes are only for redistribution (i.e. they won't be used to finance the military). Note that this idea differs from Tilly's main thesis that war makes states because of the need for these states to pay for the costs of war (although he considers this possibility in \citealp[pp. 99-103]{Tilly1990}). As we will see later, this assumption allows for tractable expressions for the effect of an increase in the risk of external conflict on political turnover and fiscal capacity, thus enabling us to identify each effect and establish a meaningful relationship between them. In addition, as will be specified below, the conditional probabilities of victory of each group are assumed to satisfy some restrictions that are consistent with a scenario where some resources could be used  to finance the military. Thus, it is possible to interpret these probabilities as affected by other resources.} However, whether $O_1$ triggers a civil war is endogenous; therefore, the ex-ante probability of political turnover (which will be key to the main result) will be endogenous.

Let $\epsilon$ denote the probability that $O_1$ gains power in period 2 if there is no civil war or external conflict. Let $\delta$ be the probability that $O_1$ wins the civil war when there is a civil war but not an external conflict. Let $\rho$ be the probability that $F$ wins the interstate war when $O_1$ triggers a civil war and there is an external conflict. Let $\mu$ be the probability that $F$ wins the interstate war when $O_1$ does not trigger a civil war and there is an external conflict. Let $\omega$ denote the probability that $O_1$ wins a civil war if at the same time $D$  is involved in an external conflict with $F$. 

I make four main assumptions that will simplify the exposition and allow the analysis to focus on interesting cases. First, I assume that the probability that $F$ wins the interstate war is larger when $O_1$ triggers a civil war:
\vspace{-0.2cm}
\small
\begin{equation}
\label{phomu}
\rho > \mu 
\end{equation}
\normalsize

Second, I assume that  $O_1$ is more likely to win a civil war if there is also an external conflict:
\vspace{-0.2cm}
\small
\begin{equation}
\label{phiowwp}
\omega > \delta 
\end{equation}
\normalsize
Note that (\ref{phomu}) and (\ref{phiowwp}) imply that internal and external conflicts reinforce each other: to secure victory, it is a good idea for both $O_1$ and $F$ to be more aggressive when their common enemy ($I_1$) is already involved in a conflict.\footnote{Of course, this does not mean that $O_1$ will always prefer to attack $I_1$ when there is an interstate conflict; the analysis will focus on the conditions under which this happens.}  

Third,  I assume that the probability that $O_1$ gains power if there is not a civil war or an external conflict is larger than $O_1$'s probability of winning the elections when there is an external conflict.\footnote{To simplify the exposition, this last probability is set to zero; thus, this assumption is equivalent to $\epsilon>0$, which  is implied by Eq. (\ref{epsilonmu}) below.} This third assumption reflects the idea that leaders are more likely to be re-elected when there is an international conflict, either because war provides them with unique opportunities to deal with their opposition \citep{ChozzaGoemans2004}, or because they engage in a ``gamble for resurrection'' \citep{DownsRocke1994}.

Fourth, I assume that the probability that $O_1$ gains power if there is no civil war or external conflict is larger than $F$'s probability of winning the interstate conflict when there is no civil war: 
\vspace{-0.2cm}
\small
\begin{equation}
\label{epsilonmu}
\epsilon > \mu 
\end{equation}
\normalsize
This assumption is made to simplify the exposition, to rule out trivial cases in which an increased risk of external conflict always prevents any investment in fiscal capacity  because it increases political instability too much.

\subsubsection*{\emph{Timing}}
\noindent The timing of the game is as follows: 
\begin{enumerate}
\itemsep0.08em 
\item[(1)] Nature decides the initial stock of fiscal capacity $\tau_1$, and $I_1\in\{A,B\}$. 
\item[(2)] $I_1$ chooses a set of period-1 policies $\{t_1,\textbf{r}_1=\{r_1^A,r_1^B,r_1^F\}\}$  and determines (through investment) the period-2 stock of fiscal capacity, $\tau_2$. 
\item[(3)] $O_1$ observes $\tau_2$ and decides whether or not to start a civil war. At the same time, an interstate conflict between $F$ and $D$ occurs with probability $\alpha$.
\vspace{-0.1cm}
\begin{itemize}[leftmargin=0.15in]
\renewcommand{\labelitemi}{$\circ$}
\item If there is no interstate conflict or civil war, $O_1$ wins the election with probability $\epsilon $ and $I_1$ remains in office with probability $1-\epsilon$. 
\item If there is no interstate conflict but there is a civil war, $O_1$ forms the new government with probability $\delta $, and $I_1$ remains in office with probability $1-\delta$. 
  \item   If there is an interstate conflict and a civil war, $O_1$ is the new government with probability $\omega+\rho\lambda$, $F$ establishes a foreign administration with probability $\rho(1-\lambda)$  and $I_1$ remains in power with probability $1-\omega-\rho$.
  
  \item If there is an interstate conflict but no civil war,  $O_1$ is the new government with probability $\mu\lambda$, $F$ establishes a foreign administration with probability $\mu(1-\lambda)$  and $I_1$ remains in power with probability $1-\mu$.
 
\end{itemize}

\item[(4)]  $I_2\in\{A,B,F\}$ chooses a set of period-2 policies $\{t_2, \textbf{r}_2=\{r_2^A,r_2^B,r_2^F\}\}$.
 \end{enumerate}


\subsubsection*{\emph{Preferences}}

\noindent The utility function of a typical member of group $J\in\{A,B\}$ in period $s$ is
\small
\begin{equation}
\label{us}
u_s^J=(1-t_s)m+r_s^J
\end{equation} \normalsize
where $m$  is an exogenous income, $t_s$ is the income tax rate and $r_s^J$ is the government transfer.  The utility function of a typical individual in group $F$ in period $s$ is
\small
\begin{equation}
\label{uf}
u_s^F= r_s^F
\end{equation}
\normalsize
where $r_s^F$ is the transfer. Note that since $F$ is a foreign group, members of $F$ are not taxed by the government of country $D$.

 
\subsubsection*{\emph{Government budget constraint}}

The income tax rate, $t_s$, is constrained by the existing fiscal capacity, $\tau_s$, such that $t_s\leq \tau_s$.  In addition,  $\tau_s$, initially set to $\tau_1$, can be augmented by non-negative investment in period 1, with increasing  and strictly convex costs $C(\tau_2-\tau_1)$, where $C_\tau(0)=0$, with $C_\tau$  denoting the partial derivative. Finally, the total population of both $D$ and $F$ is normalized to one.  Thus, the government budget constraint is\footnote{\label{difpopshares}For a more general case in which $D$ is composed of groups with different populations, the budget constraint will be $t_1m=C(\tau_2-\tau_1)+\beta^Ar_1^A+ \beta^B r_1^B+r_1^F$ in period 1 and $t_2m=\beta^Ar_2^A+ \beta^B r_2^B+r_2^F $ in period 2, where  $\beta^J$ is the population share of group $J\in\{A,B\}$. As previously mentioned, that $\beta^A=\beta^B=\vfrac{1}{2}$ is without loss of generality for the main results. Results for the general case are available upon request.}
\small
\begin{equation}
\label{BC}
\text{Budget constraint} \equiv \begin{cases}
  t_1m=C(\tau_2-\tau_1)+\vfrac{(r_1^A+ r_1^B)}{2}+r_1^F    & \text{in period } s=1 \\
  t_2m=\vfrac{(r_2^A+ r_2^B)}{2}+r_2^F     & \text{in period } s=2
\end{cases}
\end{equation}
\normalsize

\vspace{-0.5cm}
\subsubsection*{\emph{Allocation of transfers}}

In the baseline model, I assume that the opposition must receive a fixed share of the government's transfers to its own members. I distinguish between two cases: one in which the government is one of the domestic groups, $A$ or $B$, and the other in which the government is the foreign country, $F$.  For the first case, I assume that
\begin{equation}
\label{ro1}
r_s^{O_s}= \sigma^Dr_s^{I_s}
\end{equation}
where $I_s,O_s\in\{A,B\}$, and where $\sigma^{D}\in[0,1]$ denotes the fixed share of $I_s$'s transfers that must be given to the domestic opposition, $O_s$.\footnote{\label{coup}Note that for the case of $I_2=O_1$, (\ref{ro1}) implies that the period-2 government must give  $\sigma^{D}$ to the opposition \emph{regardless} of whether $O_1$ came to power through a civil war. This assumption is inconsistent with a successful revolution by the domestic opposition affecting the level of institutional cohesiveness in the country (as in \citealt{AcemogluRobinson2000, AcemogluRobinson2001}). In the online Appendix, I consider an extension of the basic model in which $\sigma^{D}=0$ when $I_2=O_1$ after a civil war. The main results hold.}   This extremely simple way of modelling the allocation of transfers, which closely follows \emph{B\&P}'s framework, tries to capture the existence of institutional arrangements that make policymakers internalize the preferences of a larger share of the population. In this respect, $\sigma^D$ can be interpreted as the level of cohesiveness of institutions. One real-world example might be the level of protection for minorities resulting from constraints on the executive (e.g. a constitutional separation of powers).  Another example might be the strength of the opposition's political representation in policy decisions, such as through proportional representation elections.

Since both interpretations of $\sigma^D$ suggest that it represents the domestic rules under which  decisions are made, it is crucial to have an intuition for where it comes from. In Section 3.4, I propose an extension of the baseline model in which $\sigma^D$ results from a bargaining process between the domestic government and opposition. In this specific context, $\sigma^D$ can be interpreted as the institutional arrangement that guarantees that in period 2, when the government is a domestic group and $I_1$'s probability of reelection is sufficiently high,  $O_1$ will receive a  large enough transfer to dissuade it from triggering a civil war in period 1.

When $F$ takes power in period 2, the opposition consists of two groups ($A$ and $B$), and it is reasonable to expect that their transfers are group-specific. In this scenario, I assume that $O_1$ will receive a larger share of $F$'s transfers.  A justification may be that  since $D$ is governed by $I_1$ when the interstate conflict takes place, $F$ might see $I_1$ as its main enemy. Thus, when there is a foreign administration, we can expect that $I_1$ will be hit the hardest. To simplify the exposition, $I_1$'s share of transfers are set to zero, i.e., $r_2^{I_1}=0\times r_2^F$. As for $I_1$, I define 
\small
\begin{equation}
\label{ro2}
r_2^{O_1}=\sigma^{F} r_2^F
\end{equation} 
\normalsize
 where $\sigma^{F}\in[0,1]$ denotes the fixed share of $F$'s transfers that must be given to $O_1$.\footnote{I assume that $\sigma^F$ is known to all the players.  However, the main results are consistent with a model in which $\sigma^F$ is known to both $O_1$ and $F$, but not to $I_1$.}
 
For an intuition about where $\sigma^{F}$ comes from, first note that $\sigma^{F}$ regulates the allocation of transfers between a foreign country occupying $D$, and a domestic group in $D$. This suggests that it may not be appropriate to think of $\sigma^{F}$ as constitutional checks and balances. To keep the model simple and tractable, I do not propose any process to derive $\sigma^{F}$; I simply assume that $\sigma^{F}$ is exogenous. A very high-level intuition is  possible: $\sigma^{F}$ may result from a previous bargaining process between $O_1$ and $F$, the outcome of which may be based on a shared trait (e.g. ethnicity, ideology, or geography). In this sense, $\sigma^{F}$ can be also understood as an ex-ante level of disunity between the government and the opposition in a country (relative to that between the opposition and a foreign country). The literature on the transnational dimensions of civil wars suggest that these kinds of factors can play an important role in explaining military decisions \citep[see][]{DavisMoore1997, Gleditsch2007, GleditschSalehyanSchultz2008, cedermangirardingleditsch2009, cedermangleditschsalehyanwucherpfennig2013, CunninghamGleditschSalehyan2011}.

\subsection{Equilibrium Policy}

I will now solve the game by backward induction. Leaving the formal derivation to Appendix A.2, here I focus on the main  equilibrium decisions. 

\subsubsection*{\emph{Civil war}}

The first main equilibrium decision is $O_1$'s decision about whether to start a civil war.  When deciding whether to trigger a civil war,  $O_1$ evaluates its expected utility based on each possible outcome. 
After computing and comparing $O_1$'s expected utilities depending on whether there is civil war or internal peace,  it is possible to show (see Appendix A.2) that $O_1$ decides to trigger a civil war when $\sigma^F>\overline{\sigma}^F$, where  
  \small
 \begin{equation}
\label{sigmaFthreshold}
\overline{\sigma}^F\equiv \frac{2\big(\alpha(\rho-\mu)(\sigma^D-\lambda)-(1-\sigma^D)(\alpha\omega  +(1-\alpha)\delta -(1-\alpha)\epsilon )\big)}{\alpha(\rho-\mu)(1-\lambda\sigma^D)+(1-\sigma^D)(\alpha\omega  +(1-\alpha)\delta -(1-\alpha)\epsilon )}
\end{equation}
 \normalsize
 provided that the denominator is positive. Note that in (\ref{sigmaFthreshold}), we have defined a threshold value for $\sigma^F$, $\overline{\sigma}^F$, such that when  $\sigma^F>\overline{\sigma}^F$, $O_1$ decides to trigger a civil war, and when $\sigma^F\leq \overline{\sigma}^F$, $O_1$ decides not to start a civil war.  
 
 From (\ref{sigmaFthreshold}), note that when $\sigma^D=0$, $\sigma^F>\overline{\sigma}^F$, so internal peace is only possible for positive levels of institutional cohesiveness (i.e for $\sigma^D>0$). When institutions are perfectly cohesive (i.e. when $\sigma^D=1$), there is no civil war (i.e. $\sigma^F<\overline{\sigma}^F$). More generally, it is possible to show that a civil war is less likely for higher levels of institutional cohesiveness (see Appendix A.2).  This is consistent with the finding that countries with many checks and balances tend to be less prone to civil wars \citep{ReynalQuerol2002DPE, ReynalQuerol2005}.
 
Also note from (\ref{sigmaFthreshold}) that a higher $\lambda$ increases  the chances of a civil war (see Appendix A.2  for a proof). So in countries where a foreign power is more likely to intervene to help the opposition (rather than to establish a foreign administration), the opposition is more prone to starting a civil war. If a larger $\lambda$ characterizes more modern forms of conflict (e.g. those occurred during the post-World War II period, as briefly discussed in footnote \ref{discussionlambda}), then this result is consistent with the increase in the number of internal conflicts in the last 60 years \citep{StrandRustadUrdalNygard2019}.

It is also possible to show from (\ref{sigmaFthreshold}) that an increased risk of an external conflict may increase or decrease the chances of a civil war. This depends on how the probability that $O_1$ gains power in period 2 changes in the absence of an external conflict: when a civil war is very attractive when there is no risk of external conflict, then an increased risk of external conflict makes a civil war less attractive (see Appendix A.2). 

\subsubsection*{\emph{Investment in fiscal capacity}}

The second main equilibrium decision is  $I_1$'s decision to invest in fiscal capacity.  To simplify notation and analysis,  I define the unconditional probability of political turnover from $I_1$'s perspective, i.e. the probability that, from $I_1$'s perspective,  $I_2\neq I_1$. Let $\phi$ denote this probability, where
\small
\begin{equation}
\label{probturnov}
\phi  =  \gamma(\alpha\omega+(1-\alpha)\delta+\alpha\rho )+(1-\gamma)(\alpha\mu+(1-\alpha)\epsilon)
\end{equation}
\normalsize
where $\gamma=\gamma(\sigma^F)$ is the indicator function taking the value of 1 if $\sigma^F>\overline{\sigma}^F$ and 0 otherwise.\footnote{When $\sigma^F$ is unknown to $I_1$, $\gamma$ denotes $I_1$'s beliefs about $\sigma^F$.} 

\medskip
After computing $I_1$'s first- and second-period indirect utilities, as well as $I_1$'s expected utility as seen from period 1, it is possible to show (see Appendix A.2, proof of Prop. 1)  that the $\tau_2$ that maximizes $I_1$'s expected utility is
\small
\begin{equation}
\label{tau2fin}
\tau^*_2 = C_\tau^{-1}\big(m\big[-\phi(1-\sigma^D)-\alpha(\gamma\rho+(1-\gamma) \mu)(1-\lambda)\sigma^D+\vfrac{(1-\sigma^D)}{2}\big]\big) +\tau_1.
\end{equation}
\normalsize

\medskip
Since this paper focuses on how the risk of an external conflict affects investments in fiscal capacity, I look at $\frac{\partial \tau^*_2}{\partial \alpha}$.  However, given that the expression for $\tau^*_2$ in (\ref{tau2fin}) depends  on political turnover (as perceived by $I_1$), and that political turnover will play a crucial role in the interpretation of the main results, I start by identifying the conditions under which an increased risk of an external conflict increases or decreases  $\phi$. 

  \begin{prop1} Consider the above-described game. Then, there is a unique threshold for $\sigma^F$, denoted by $\overline{\sigma}^F$ and given by (\ref{sigmaFthreshold}), such that in equilibrium: 
  \begin{itemize} \itemsep-0.1em 
      \item[(1.A)]  If $\sigma^F>\overline{\sigma}^F$, an increased risk of an external conflict increases  $\phi$. 
       \item[(1.B)] If $\sigma^F\leq\overline{\sigma}^F$, an increased risk of an external conflict decreases  $\phi$. 
\end{itemize}
 \end{prop1}
 \begin{proof}
See Appendix A.2.
\end{proof} 

Prop. 1 states that an increased risk of an external conflict can either increase or decrease political turnover (as perceived by the period-1 incumbent, $I_1$), and that this depends on the size of the transfers that the period-1 opposition ($O_1$) expects to receive if there is a foreign administration in period 2 (which, as previously mentioned, can be understood as a measure of the ex-ante level of disunity between the government and the opposition). Importantly, Prop. 1 shows that a higher risk of external conflict can decrease  political turnover (Prop. 1.B) and this happens because the higher risk of external conflict makes $O_1$ less belligerent.\footnote{This result can be interpreted as specifying how a sense of national identity that creates common interests can be actively fostered: when the ex-ante level of disunity between the government and the opposition is sufficiently small, a foreign invasion is more likely to be perceived as a common threat, so when that threat increases, national cohesiveness, reflected in a lower probability of civil war, also increases. This interpretation is consistent with the literature on endogenous social identity \cite[for example, see][]{Shayo2009}, providing an alternative mechanism through which a sense of identity and belonging to a polity can be reinforced.}  In this scenario, $I_1$ is strengthened domestically when the reduction in $O_1$'s  belligerence outweighs the higher probability that $F$ wins power.

\medskip

 Now I examine how an increased risk of external conflict  affects investments in fiscal capacity. The following proposition summarizes the result, which constitutes the first main result of the paper. 

  \begin{prop2} Consider the above-described game. Then, there is a unique threshold for $\sigma^F$, denoted by $\overline{\sigma}^F$ and given by (\ref{sigmaFthreshold}), such that in equilibrium: 
  \begin{itemize} \itemsep-0.1em 
      \item[(2.A)]  If $\sigma^F>\overline{\sigma}^F$, then an increase in the risk of external conflict decreases investments in fiscal capacity. 
       \item[(2.B)] If $\sigma^F\leq\overline{\sigma}^F$, then: 
       		\begin{itemize}\itemsep-0.1em 
  			\item[(2.B.1)] If $(\epsilon-\mu)>\sigma^D(\epsilon-\lambda\mu)$,  an increase in the risk of external conflict increases investments in fiscal capacity.
			\item[(2.B.2)] If $(\epsilon-\mu)=\sigma^D(\epsilon-\lambda\mu)$,  an increase in the risk of external conflict does not affect investments in fiscal capacity. 
 		 	\item[(2.B.3)] If $(\epsilon-\mu)<\sigma^D(\epsilon-\lambda\mu)$,  an increase in the risk of external conflict decreases investments in fiscal capacity. 
		\end{itemize}
\end{itemize}
 \end{prop2}
 \vspace{-0.4cm}
 \begin{proof}
See Appendix A.2.
\end{proof} 
 
Prop. 2 states that an increase in the risk of external conflict can either increase or decrease the government's incentives to invest in fiscal capacity; this depends on  the size of the transfers that the period 1 opposition ($O_1$)  expects to receive if there is a foreign administration. As previously mentioned, the intuition for this result crucially depends on how an increased risk of external conflict affects political turnover,  and, specifically, on whether a heightened risk of an external conflict increases the chance that the incumbent in period 1 ($I_1$) gets support from $O_1$, so that $I_1$ strengthens its domestic support. 

Prop. 2.A considers the case where $I_1$ does not expect to receive any support from $O_1$ because $O_1$ is expected to receive large transfers if there is a foreign administration. In this scenario, $O_1$ sees an opportunity to weaken $I_1$ by triggering an internal conflict. The consequence is that $I_1$ expects a higher likelihood of political turnover (Prop. 1.A), which reduces its incentives to increase period-2 fiscal capacity.\footnote{This result is consistent with the Mexican-American War (discussed in Section \ref{cases}) and the Shaba wars (discussed in Appendix A.1): the absence of any strong sense of unity could have increased the risk of a coup, which in turn could have given the Mexican and DRC governments incentives to raise revenues by means other than direct taxation.} 

Prop. 2.B considers the case where $O_1$ expects to receive small transfers if there is a foreign administration. In this scenario, $I_1$ expects to receive support from $O_1$, which increases  $I_1$'s expectation of  staying in period 2 (Prop. 1.B).  This proposition identifies a new condition under which the lower probability of turnover, as seen by $I_1$, provides $I_1$ with enough incentives to increase period-2 fiscal capacity.\footnote{This result is consistent with Guerrero's presidency (discussed in Section \ref{cases}): in the early months of this presidency, there was a high risk of reconquest by Spain, and most Mexicans sided with the government in strongly opposing Spain (recall that the Mexican government had massively expelled people who may have eventually sided with Spain). This unity may have decreased the risk of political turnover, giving Guerrero incentives to insist, successfully, on the introduction of a national income tax.} 

The explanation for the new condition is as follows. First, note that it relates  two terms: first, a term that represents the magnitude of the increase in  $I_1$'s expectation of staying in power in period 2, ($\epsilon-\mu$), which results from an increased risk of external conflict; and second, a term that represents the magnitude of the decrease in the chance that $O_1$ is in power in period 2, $(\epsilon-\lambda\mu)$, multiplied by the level of cohesiveness of institutions, $\sigma^D$.\footnote{Note that the increased chance that $I_1$ stays in power is smaller than the decreased chance that  $O_1$ is in power in period 2  (i.e,  $(\epsilon-\mu)<(\epsilon-\lambda\mu)$); thus, it matters that $(\epsilon-\lambda\mu)$ is multiplied by $\sigma^D$.} When the first term is greater than the second term, an increased risk of external conflict increases investment in fiscal capacity (Prop. 2.B.1), and when the first term is smaller than (or equal to) the second term, an increased risk of external conflict does not increase investment in fiscal capacity (Props. 2.B.2 and 2.B.3). 

The intuition for when the first term is greater (or smaller) than the second term crucially depends on $\sigma^D$. To see this, note that the only benefit to $I_1$ of a higher probability of staying in power in period 2 (which we know happens because of Prop. 1.B)  is that second-period rent extraction becomes more likely. The costs include a higher probability that $F$ extracts $I_1$'s period-2 rents (because a foreign administration becomes more likely), and a lower probability that $I_1$ receives transfers from $O_1$ in period 2 (because the chance that $O_1$ is in power in period 2 decreases). Importantly, both  the amount of rents $I_1$ expects to capture if it stays in power, and the transfers $I_1$ expects to receive if $O_1$ is in power in period 2, inversely depend on the cohesiveness of institutions: the weaker the government institutions, the more rents can be captured by the incumbent and the smaller the share that must be transferred to the opposition.  So, if $I_1$ expects that rent extraction is more likely, and the value of the rents that it can capture is relatively high,\footnote{But not too high, because if so, $O_1$ would prefer to trigger a civil war (as in Prop. 2.A).} then $I_1$'s marginal benefit from investing more in fiscal capacity can be relatively high. In addition, if $I_1$ expects that $O_1$ will not be in power in period 2, then the fact that $O_1$ would be able to capture more rents if it were in power in period 2 becomes less costly. In this scenario, investment in period-2 fiscal capacity is more likely, as stated in Proposition 2.B.1.

The intuition for Propositions 2.B.2 and 2.B.3 is similar: the stronger the government institutions, the less rents can be captured by the incumbent and the greater the share that must be transferred to the opposition. Thus, if $I_1$ expects that rent extraction is more likely, but the value of the rents that it can capture is relatively low,  then $I_1$'s marginal benefit from investing more in fiscal capacity can be relatively low. In addition, if $I_1$ expects that $O_1$ is not in power in period 2, then $O_1$ being able to capture less rents if it were in power in period 2 is more costly.

  \begin{figure}
\begin{center}
\caption{Parameters and different types of outcomes}\label{emsigma}
\vspace{-0.4cm}
\resizebox{11cm}{6.2cm}{\includegraphics[width=4in]{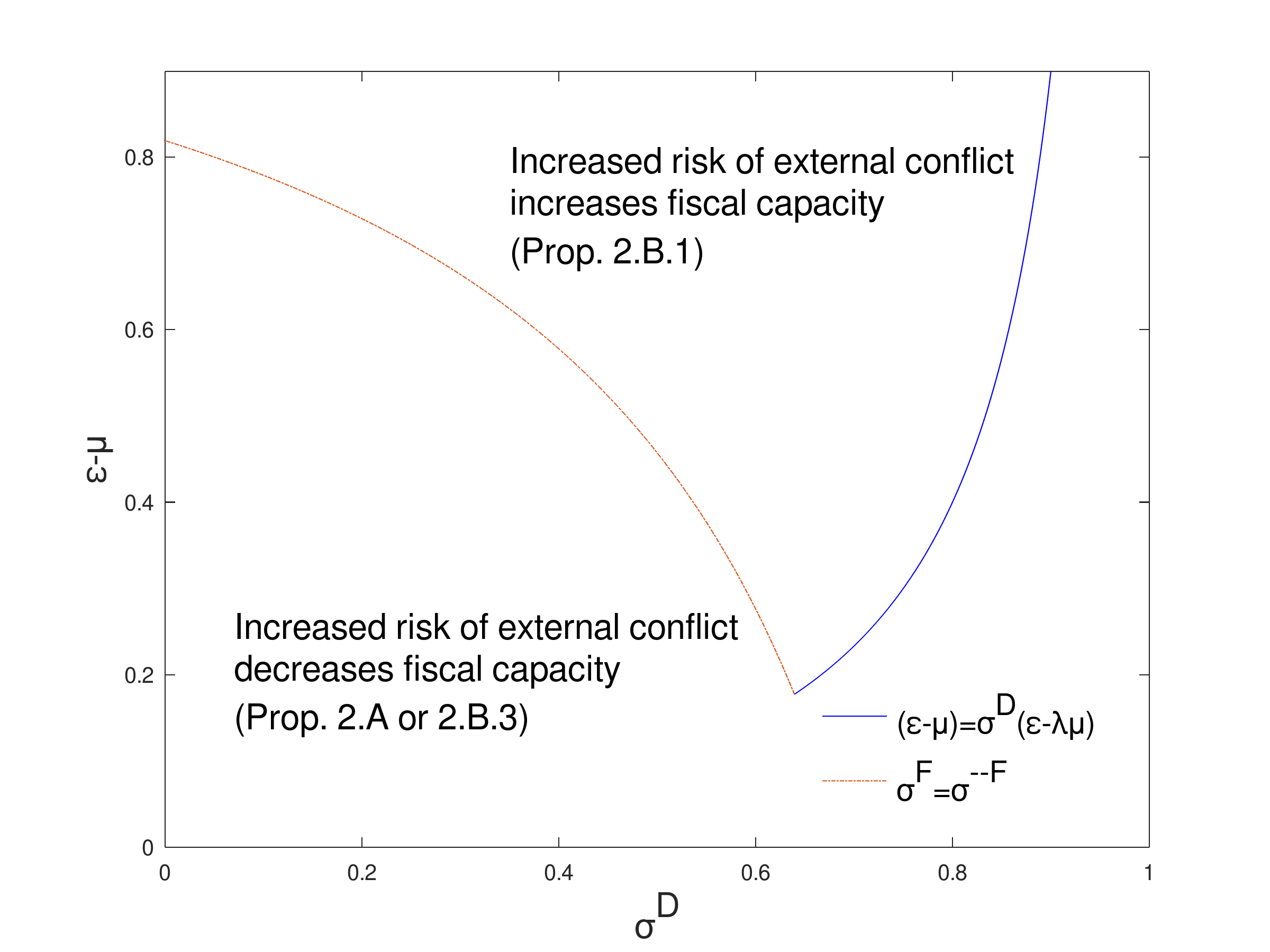}}
\subcaption*{\footnotesize{Other simulation parameter values: $\alpha=0.5$, $\rho=0.5$, $\omega=0.5$, $\delta=0.4$, $\mu=0.1$, $\sigma^F=0.1$, $\lambda=0$}}
\end{center}
\end{figure}

 Figure \ref{emsigma} illustrates the conditions under which  an increased risk of external conflict increases  or decreases investments in fiscal capacity for combinations of the key parameters mentioned in Prop. 2 ($\sigma^D$ and $\epsilon-\mu$). The figure illustrates the intuition proposed in the last paragraphs for the conditions for an increase in fiscal capacity to result from an increased risk of external conflict: it requires i) an intermediate level of institutional cohesiveness, and ii) a sufficiently high increase in  $I_1$'s expectation of staying in power in period 2.
 
 An additional result can be found by looking at how the conditions in Props. 2.B.1 to 2.B.3 depend on whether  a foreign intervention results  in the strengthening of the domestic opposition rather than in a foreign administration (i.e. how they depend on $\lambda$). In this respect, note that the condition in 2.B.1 (i.e. $(\epsilon-\mu)>\sigma^D(\epsilon-\lambda\mu)$) is more likely to be satisfied when a foreign intervention strengthens the domestic opposition (i.e. when $\lambda$ is big), and less likely when it results  in a foreign administration (i.e. when $\lambda$ is small). The intuition is straightforward, and sheds additional light on the costs associated with interstate conflicts that result in a foreign administration: since $I_1$'s worst-case scenario is a foreign administration, state-building is more likely when a direct foreign intervention primarily seeks a new distribution of power between the domestic actors.
 
 \medskip

Propositions 1.B and 2.B and the explanation proposed in the last paragraph reveal an interesting characteristic of the state-building process: although political stability crucially affects incentives for state-building, more political stability does not always imply more state-building. This is summarized in the following proposition, which constitutes the second main result of the paper.

    \begin{prop3} Consider the above-described game. Then, there is a unique threshold for $\sigma^F$, denoted by $\overline{\sigma}^F$ and given by (\ref{sigmaFthreshold}), such that in equilibrium: 
    \begin{itemize} \itemsep0em 
  \item[(3.A)] If $\sigma^F>\overline{\sigma}^F$, an increased risk of external conflict  implies an increase in $\phi$ and decreased investment in fiscal capacity. 
  \item[(3.B)] If $\sigma^F\leq\overline{\sigma}^F$, then: 
       		\begin{itemize}\itemsep-0.1em 
  			\item[(3.B.1)] If $(\epsilon-\mu)>\sigma^D(\epsilon-\lambda\mu)$,  an increase in the risk of external conflict implies a decrease in $\phi$ and increased investment in fiscal capacity. 
			\item[(3.B.2)] If $(\epsilon-\mu)\leq\sigma^D(\epsilon-\lambda\mu)$,  an increase in the risk of external conflict implies a decrease in $\phi$ but not increased investment in fiscal capacity. 
		\end{itemize}
  \end{itemize}
    \end{prop3}
    \begin{proof}
See Appendix A.2. 
\end{proof}

Prop. 3 is consistent with the important but still scarce literature on the relationship between greater political stability and weaker public policies, particularly an interesting result by \cite{AcemogluGolosovTsyvinski2011}: greater stability might harm the party in power because when power finally shifts, the new government is more likely to remain in power, keeping a bigger share of rents and leaving less to the former government. Prop. 3.B contributes to this idea by providing a novel mechanism through which more political stability might be associated with worse outcomes, and in which the origin of the greater political stability is key: when greater political stability results from an increased risk of external conflict,  the higher chance that the government loses power to the foreign country means that the costs associated with more fiscal capacity in period 2 are higher,  so only non-cohesive institutions can give the government enough incentives to invest in fiscal capacity.\footnote{Note that this result cannot be deduced from \emph{B\&P}'s model, since for them an increase in the risk of external conflict always increases the value of a common-interest public good, so greater political stability that results from a greater risk of external conflict always implies more incentives to invest in fiscal capacity.} 


\normalsize
\subsection{Endogenous political institutions}

In this section, I expand the choices in the baseline model to include the level of cohesiveness of institutions, $\sigma^D$.  I consider a very simple way in which this parameter can change: it results from a bargaining process between the period-1 government and the opposition. 

In the new scenario, the level of cohesiveness of institutions is specific to each period, so we have $\sigma^D_1$ for period 1 and $\sigma^D_2$ for period 2; $\sigma^F$ is still exogenous. 

The analysis focuses on $\sigma^D_2$. Importantly, I assume that $\sigma^D_2$ represents a constitutional change made one period ahead, and that an agreement, if reached, deters the domestic opposition from starting a civil war and ensures the period-2 government does not use its power to modify $\sigma^D_2$.\footnote{The new timing of the game differs from the baseline model in that at stage 2, the second-period political institutions $\sigma^D_2$ are chosen. At stage 1, nature decides the initial level of cohesiveness of institutions, $\sigma^D_1$. In this scenario, I introduce two new assumptions: i) $\vfrac{1}{2}\geq\epsilon$ and ii) $\epsilon=\delta$. The first assumption is plausible in scenarios in which each group makes up half of the population (which we assume) and where there is some incumbent advantage. The second assumption simplifies the exposition.} This assumption is weaker than in the baseline model, and although it could be still regarded as a strong assumption,  it is not uncommon in the literature on endogenous institutions.\footnote{In addition to \emph{B\&P}, see \cite{AghionAlesinaTrebbi2004}, \cite{TicchiVindigni2010} and \cite{AcemogluRobinsonTorvik2013revstud}. 
} 
A possible intuition is an agreement that includes, on the one hand,  significant transaction costs in unilaterally changing future key political institutions (for instance, through entrenched clauses that impose constitutional limits on the government) and, on the other hand, refraining from the use of violence for political purposes (for example, through the disarmament of $O_1$).

The bargaining process is modeled as a very simple ``divide-the-dollar'' game. First, the period-1 government, $I_1$, offers a certain value of $\sigma^D_2$ to the opposition, $O_1$. $O_1$ then chooses whether to accept the offer, taking into account that if it accepts, the proposed value of $\sigma^D_2$ will determine both $O_1$'s and $I_1$'s transfers in the second period provided that one of the domestic groups is in power in period 2. If $O_1$ accepts the offer, the constitution will be implemented in period 2 if there is no foreign administration, and each domestic group will receive a transfer determined by (\ref{ro1}).  If $O_1$ rejects the offer, $\sigma^D_2$ is set at its lowest level, i.e. $\sigma^D_2=0$,  which can be interpreted as a situation in which $\sigma^D_1=0$, and the status quo is maintained. Finally,  if $O_1$ accepts the offer, it commits to not starting a civil war.\footnote{The opposition takes into account its commitment not to start a civil war when determining whether or not to accept the offer. As previously mentioned, the commitment can be interpreted as resulting from a peace deal in which the opposition may have turned over its weapons in exchange for a certain level of cohesiveness of institutions protected by a set of entrenched clauses.}

To solve this variation of the model, I proceed by backward induction, first considering $O_1$'s choice at stage 2, and then the conditions under which it is optimal for $I_1$ to offer something that $O_1$ will accept. The following proposition summarizes the main result:

\begin{prop4}  Consider the above-described game with $\sigma^D_2$ resulting from a bargaining process between the period-1 government and the opposition. Consider the inequalities 
\small
\begin{equation}
\label{condprop40}
\alpha\omega+\alpha\rho\lambda+\alpha\mu(1-\lambda)/2+(1-\alpha)\delta+\alpha(\rho-\mu)(1- \lambda)\sigma^F/(2+\sigma^F)\leq \vfrac{1}{2}
\end{equation}
\begin{equation}
\label{condprop4}
(1-\alpha)\epsilon+\alpha\mu(1+\lambda)/2<\alpha\omega+\alpha\rho\lambda+\alpha\mu(1-\lambda)/2+(1-\alpha)\delta+\alpha(\rho-\mu)(1- \lambda)\sigma^F/(2+\sigma^F)
\end{equation}
     \vspace{-0.1cm}
\normalsize
Then, 
\begin{itemize}\itemsep-0.1em 
  \item[(4.A)]   if (\ref{condprop40}) and (\ref{condprop4}) are satisfied, the period-1 government will offer $\sigma^{D^*}_2>0$, and the opposition will accept it, where
 \small
 \begin{equation}
\label{condprop4a}
\sigma^{D^*}_2= \frac{\alpha(\omega+(\rho-\mu)\lambda)+\alpha(\rho-\mu)(1- \lambda)\sigma^F/(2+\sigma^F)}{1-\alpha(\omega+\mu+\rho\lambda)-2(1-\alpha)\epsilon-\alpha(\rho-\mu)(1- \lambda)\sigma^F/(2+\sigma^F)}
\end{equation}
\normalsize
     \vspace{-0.4cm}
  \item[(4.B)]  if  (\ref{condprop40}) is satisfied but  (\ref{condprop4}) is not satisfied,
 the period-1 government will offer $\sigma^{D^*}_2=0$ and the opposition will accept  the offer. 
  \item[(4.C)]   if  (\ref{condprop40}) is not satisfied, the period-1 government will offer $\sigma^{D^*}_2=0$, and the opposition will reject the offer. 
\end{itemize}

\end{prop4}  
 
\begin{proof}
See Appendix A.2. 
\end{proof}

The condition in (\ref{condprop40}) guarantees that $I_1$ will make an offer that  $O_1$ will accept, which implies that there won't be an internal conflict. When (\ref{condprop40}) holds,  (\ref{condprop4}) determines whether the level of cohesiveness of institutions is greater than zero. 

The intuition for these results is the following. First, note that when (\ref{condprop40}) holds, the probability that $O_1$ takes power through a civil war  and $O_1$'s marginal benefit in the event of a foreign conflict cannot be very high. This implies that $O_1$ will accept the offer because these expressions are directly proportional to $O_1$'s bargaining power, so when they are sufficiently small, so is this power.  When (\ref{condprop4}) also holds (Prop. 4.A), the condition can also be related to $O_1$'s bargaining power, but now implies that it cannot be extremely low. Since in this scenario the probability of political turnover is small, $O_1$ expects that  $I_1$ is in power in period 2, so $O_1$'s bargaining power implies that $I_1$ has to propose a positive --- but not necessarily high --- level of checks and balances for period 2. When (\ref{condprop4}) does not hold (Prop. 4.B),  $O_1$'s bargaining power is extremely low, so $O_1$ will accept any proposal, including $I_1$'s  preferred alternative,  where the level of institutional cohesiveness is null.

Note that the equilibrium level of cohesiveness of institutions in (\ref{condprop4a}) increases with $\sigma^F$ and $\alpha$.\footnote{
That $ \vfrac{\partial \sigma^{D^*}_2}{\partial \sigma^F}>0$ follows directly from the fact that $\vfrac{\sigma^F}{(2+\sigma^F)}$ is strictly increasing in $\sigma^F$. As for $ \vfrac{\partial \sigma^{D^*}_2}{\partial \alpha}>0$, see Appendix A.2 (proof of Proposition 5).} The intuition is similar to that previously mentioned: having closer links with $F$ and a higher risk of an attack by $F$  increases $O_1$'s reservation utility, so $I_1$ has to propose more checks and balances in period 2 to give $O_1$ sufficient incentive to accept the offer.  

\medskip

As for how the risk of external conflict affects investment in fiscal capacity, note that Prop. 1 may not apply  to this scenario because now $\sigma^{D^*}_2$ can increase in $\alpha$. The following proposition  summarizes this effect:
\black
 
\begin{prop5}  Consider the above-described game with $\sigma^D_2$ resulting from a bargaining process between the period-1 government and the opposition.  Then: 
\begin{itemize}\itemsep-0.1em 
    \item[(5.A)] if (\ref{condprop40}) and (\ref{condprop4}) are satisfied, an increase in the risk of an external conflict does not increase investment in fiscal capacity. 
    \item[(5.B)]  if  (\ref{condprop40}) is satisfied but  (\ref{condprop4}) is not satisfied, and if $(\epsilon-\mu)>\sigma_2^{D^*}(\epsilon-\lambda\mu)$, an increase in the risk of external conflict increases investment in fiscal capacity. 
        \item[(5.C)]  if (\ref{condprop40}) is not satisfied, an increase in the risk of external conflict decreases investment in fiscal capacity. 
\end{itemize}
\end{prop5}   
\begin{proof}
See Appendix A.2. 
\end{proof}

The intuition for Prop. 5.A is the following. First, note that a higher risk of external conflict improves $O_1$'s outside option, since it makes it more likely that  $O_1$ takes power in the event of a civil war. Second, note that the increase in $O_1$'s reservation utility makes an agreement more costly for $I_1$ in terms of having to accept more checks and balances in period 2. Third, note that if an agreement is reached, there won't be  civil war and $I_1$'s probability of re-election will increase in the event of an external conflict.  In this scenario, Prop. 5.A states that $I_1$  has no incentive to invest in fiscal capacity because the costs associated with its decreased capacity to extract rents in period 2 are now too high.\footnote{For the same reason, when in equilibrium institutions are expected to be very cohesive, i.e. when $\sigma^{D^*}_2=1$, $I_1$ does not have an incentive to increase fiscal capacity either. We also don't expect an increase in investment in fiscal capacity when there is no agreement and a civil war is expected to occur: the increased likelihood of turnover is so big that it outweighs the potential for rent extraction.} 

Importantly, note that Prop. 5.A holds even though when an agreement is reached, $I_1$ expects a decrease in political turnover. This allows us to ask about the role of political turnover in this new result.  In this respect, note that the result in Prop. 5.A is consistent with  the most important result in Prop. 3: a decrease in political turnover  may not result in increased investment in fiscal capacity when the decrease in political turnover is a consequence of an increased risk of external conflict. The intuition here is the same; the only difference is that now the level of institutional cohesiveness is endogenous. So, in some sense, Prop. 5.A generalizes Prop. 3.B.\footnote{However, Props. 4 and 5 differ with Prop. 3 in one aspect: an increased risk of external conflict also decreases the chance that (\ref{condprop40}) is satisfied. To see this, differentiate the left side of (\ref{condprop40}) with respect to $\alpha$ and note that by (\ref{phiowwp}), the resulting expression is always positive, so a higher risk of external threat makes it less likely that (\ref{condprop40}) is satisfied.} 

Proposition 5.B shows that an increased risk of external conflict can increase investments in fiscal capacity. This occurs when $O_1$'s bargaining power is very weak, so $I_1$ can get support from $O_1$ even though the level of cohesiveness of institutions  is very low and $I_1$ expects to be re-elected. This result sheds light on what adds the variation proposed in this section to Prop. 2. Note that when the level of cohesiveness of institutions results from a bargaining process between $I_1$ and $O_1$,  an increased risk of external conflict implies less investment in fiscal capacity because $I_1$, through a decrease in its capacity to extract rents in period 2, internalizes the costs associated with $O_1$'s support being more difficult to obtain. Crucially, note that the internalization of costs through a higher  level of institutional cohesiveness only occurs when the equilibrium level of cohesiveness of institutions is not extremely low. When it is, and $O_1$'s  bargaining power is ex ante extremely low, as it is the case in Prop. 5.B; the equilibrium level of cohesiveness of institutions is not affected by an increased risk of external conflict; and changes to investment in fiscal capacity only depend on the relationships between the other parameters. 

\section{Conclusion}
\label{conclusion}

Most of the recent literature on state capacity assumes that threats from external enemies generate common interests among groups in society, leading to larger investments in state capacity. In addition, external and internal disputes are viewed as being independent of each other. However, a large number of cases and some cross-country correlations suggest that this might not be the case. In particular, this assumption seems inconsistent with state-building processes during post-independence and Cold War periods in several countries in Latin America, sub-Saharan Africa and southeast Asia. 

Motivated by these observations, this paper develops a model where interstate conflicts and civil wars can be related, and in which interstate wars are not always a common-interest public good. In the model, a government faces two threats: one from a foreign country, and another from a domestic opposition. Crucially, the model allows for an opposition and government that place different values on a potential victory by the foreign country. A first main result establishes that in equilibrium, the risk of external conflict contributes to fiscal capacity only when the share of transfers that the internal opposition expects if it supports the government is sufficiently large relative to what it expects from a foreign administration. 

The interplay between external threats, political turnover and investment in fiscal capacity leads to a second important result: while a greater likelihood of political turnover translates into less fiscal capacity, a lower probability of political turnover does not necessarily imply more state-building. The link between political turnover and state-building depends on the cohesiveness of institutions and on the fact that an increased risk of external conflict reduces the likelihood of political turnover. If institutions are sufficiently cohesive, the gains from political stability might not outweigh the more probable losses resulting from a foreign administration, which means that decreased political turnover could actually lead to less state-building. 

An extension of the baseline model that endogenizes the level of cohesiveness of institutions is also proposed. In this extension, the cohesiveness of institutions results from a constitutional stage in which the government and opposition play a simple  ``divide-the-dollar'' game.  The results in this extension are consistent with those in baseline model. 

I view this paper as a first step towards a more systematic analysis of the process of state-building when countries experience different but related conflicts. I develop a simple model to study a specific aspect of this process: the construction of two specific institutions in a country consisting of two domestic actors that face an exogenous external threat. Many aspects of this process have been left out, such as the source of the external threat and how the losing domestic actor and former incumbent behaves when  the foreign country wins the interstate conflict. These are important and exciting areas for future research.

\hbox {} \newpage
\appendix
\section{Appendix}
\setcounter{equation}{0}
\renewcommand\theequation{A.\arabic{equation}}
\small
\subsection{Additional Case-Based Evidence: The Shaba Wars}

This appendix briefly discusses a case study from the Democratic Republic of the Congo (DRC) during Mobutu Sese Seko's rule. Like the Mexican case studies discussed in Section \ref{cases}, the DRC case study illustrates how the risk of a foreign intervention, and government and opposition responses to that risk, explain important aspects of the state-building process. 

\smallskip

DRC's early years as an independent country were chaotic. In 1965, a new regime took power led by President Mobutu, which sought to distance itself from the previous regime. Mobutu's regime would last for nearly 32 years. It progressed through various phases, all having two common elements: the occurrence of several conflicts closely tied to the Cold War, and patrimonial predatory practices with almost no effort toward building strong economic institutions (\citealp[chs. ~2, 10-12.]{YoungTurner1985}; \citealp[ch. ~5]{NzongolaNtalaja2002}).

I will now discuss a period characterized by two wars, known as the Shaba wars, in more detail. The Shaba wars occurred in 1977 and 1978, after a decade during which Mobutu had been ``relatively successful in reuniting most of the country, and ending previous disorder"  \cite[p. 733]{Young1984}. The wars were triggered by an invasion of the Shaba province (today named Katanga) by rebels based in Angola and led by former Katangan exiled gendarmes.\footnote{The term ``invasion'' is debated as the attacks were carried out by people of Cogolese descent, but it is plausible that Mobutu viewed the attacks as a foreign intervention.} The militia sought to conquer the region and remove Mobutu from power  \cite[pp. 71-75]{NdikumanaEmizet2005}. These wars coincided with a peak in US and Soviet Union interventionism \citep[p. 4]{Westad2005}.\footnote{In particular, they occurred when there was a significant risk of Soviet intervention following Soviet involvement in  Angola \cite[pp. 376-378]{YoungTurner1985}. Even though the Soviet Union desired the overthrow of Mobutu's regime and the militia used Soviet arms \cite[p. 423]{YoungTurner1985}, there seems to be no evidence of direct Soviet involvement in the invasion (\citealt[p. 75]{YoungTurner1985} and \citealt{Larmer2013}).}  
 
 The Shaba invasions were a major challenge to Mobutu's rule \cite[p. 71]{NdikumanaEmizet2005},  so much so that ``his dominion had nearly collapsed twice with the Shaba interventions"  \citep[p. 33]{Hesselbein2007}.  Importantly, as in the case of the Mexican-American war,  the invaders were ``warmly received by many'' and ``the demoralization and incapacity of the security forces were plainly evident to all'' \cite[p. 75]{YoungTurner1985}.\footnote{Mobutu was  aware of this division and lack of commitment when he declared in a major address that this crisis had revealed the betrayal of many high-placed cadres, ``who expressed doubts, began to distance themselves, kept one foot in each camp, hesitated to wear their party buttons, and privately predicted the downfall of the regime" \citep[p. 74]{YoungTurner1985}.} Following the invasion, which Mobutu fought off with the aid of both France and the US, the rebel forces  withdrew with their numbers largely intact, and it is plausible that the rebels  ``continued to pose a threat because of their ability to move amongst the civilian Zairian population'' (\citealt{Larmer2013}). 

This period was also characterized by the establishment of a complex patrimonial system in which economic resources and key offices were exchanged for personal loyalty and service to the president \cite[ch. 6]{YoungTurner1985}. Besides ending a decade of relative order and unity, Crawford Young sees the period as a turning point at which Mobutu began to appear predatory:

\begin{quote}
Until that point ... [Mobutu] appeared to cling to the illusion that he could at once realize the grandiose dreams for the country that he ceaselessly proclaimed ... past this point, personal rule became what Juan Linz, borrowing a Weberian term, labels `sultanism', whereby `personalistic and particularistic use of power for essentially private ends of the ruler and his collaborators makes the country essentially like a huge domain \cite[p. 188]{Young2012}.
 \end{quote}
 And this turn was enigmatic because, as \cite{Ogunbadejo1979} put it,
\begin{quote}
if he [Mobutu]  had really wanted to, Mobutu could have mobilized the government institution to cement national unity. Instead, he built a highly autocratic structure of government around himself and took control of every decision from the publication of a book to the granting of an import license, giving his ministers less and less responsibility. \citep[p. 33]{Ogunbadejo1979}. 
 \end{quote}

So why did Mobutu choose an autocratic structure of government precisely at the moment of greater political instability? A possible explanation is that it was precisely because of the political instability, which substantially increased during this period, as the ``parliament refused to agree to the
budget ... students started to protest ... and riots in Kinshasa were such that the regime used the troops to `pacify' them" \citep[p. 32]{Hesselbein2007}. And, as a consequence of this significant internal pressure,  ``the previous elite consensus did not hold ... [so that] ... members of the elite turned more and more to neo-patrimonial, ethnically or regionally based politics, while Mobutu presided as the top patron over these fragmenting institutions'' \citep[p. 33]{Hesselbein2007}. 

Although the establishment of Mobutu's patrimonial system coincided with the beginning of the DRC's relentless economic decline, the patrimonial system seems to have fuelled the decline by undermining any alternative policy that could have built economic institutions. The DRC faced a grave economic and financial crisis, and Mobutu sought relief from international financial institutions. These institutions made their relief conditional on the DRC implementing measures that could have strengthened government institutions: ``reduce corruption, rationalize and control expenditures, increase tax revenues, limit imports, boost production, improve the transportation infrastructure, eliminate arrears on interest payments, ensure that principal payments were made on time, and improve financial management and economic planning" \cite[p. xiv]{MeditzMerrill1994}, However, as Meditz and Merrill observe:

\begin{quote}
The thorough implementation of changes and reforms required by the World Bank, the IMF, and other Western donors was perceived as a threat to the very basis of the elite's power-access to and free use of the nation's resources. If the president were to execute effectively the reforms his foreign partners demanded, the heart of his authority: complete personal discretion and the fiscal privileges and corruption that bound the system together, would be undermined. As a result, Mobutu and the political elite used their control of government institutions to sabotage economic change \cite[p. 146]{MeditzMerrill1994}. 
 \end{quote}

In the end, Mobutu obtained resources by ``manipulating [the regime's] donors' economic interests against one another and by exploiting foreign anxieties about the instability that might result from a collapse of the regime" \cite[p. 146]{MeditzMerrill1994}. Thus, despite the high risk of an external conflict, the existence of a closely related domestic conflict, which coincided with the establishment of an extreme form of patrimonialism, blocked any effort to build economic institutions \cite[pp. 76-77]{KaiserWolters2012}. 

\smallskip

As  was the case with the Mexican-American war, this additional case study illustrates a mechanism through which the risk of external conflict can reduce incentives to invest in fiscal capacity: the risk of a victory by the insurgents seems to have reduced the chances of state-building reforms insofar as they made a collapse of Mobutu's regime more likely. This instability seems to have been key in the establishment of a patrimonial regime bent on holding onto power by any means possible.


\newpage 
\subsection{Proofs of the Results}


\small
\begin{proof}[\textbf{Derivation of threshold in Eq. (\ref{sigmaFthreshold})}]  

 I start by studying the optimal policy chosen by the government in period 2 (stage 4). There are two possible cases: one in which the government is one of the domestic groups (i.e. $I_2\in\{A,B\}$), and the other in which the government is the foreign power (i.e. $I_2=F$).   For $I_2\in\{A,B\}$, $I_2$ chooses $\{t_2, \textbf{r}_2=\{r_2^A,r_2^B,r_2^F\}\}$ to maximize (\ref{us}),  subject to (\ref{BC}), (\ref{ro1}) and $t_2\leq \tau_2$.
It is easy to see that we have a corner solution with $t_2=\tau_2$, $r_2^F=0$, $r_2^{O_2}= \sigma^Dr_2^{I_2}$ and  $r_2^{I_2}=\vfrac{2\tau_2m}{(1+\sigma^D)}$.  For  $I_2=F$, $F$ chooses $\{t_2,  \textbf{r}_2=\{r_2^A,r_2^B,r_2^F\}\}$ to maximize (\ref{uf}) subject to $t_2\leq \tau_2$, (\ref{BC}) and (\ref{ro2}). In this case we have that $t_2=\tau_2$,  $r_2^{I_1}=0$, $r_2^{O_1}=\sigma^Fr_2^F$, and $ r_2^F=\vfrac{2\tau_2 m}{(2+\sigma^F)}$. 

\medskip

I now study $O_1$'s decision about whether to start a civil war.  Let $W^{O_1}(\tau_2|I_2=K)$ be  $O_1$'s period-2 indirect utility if group $J\in\{I_1,O_1,F\}$ is the government in period 2.

\smallskip
 For $I_2\in\{A,B\}$, replacing $r_2^{I_2}=\vfrac{2\tau_2m}{(1+\sigma^D)}$ and $r_2^{O_2}= \sigma^D r_2^{I_2}$ in (\ref{us}), we have that $O_1$'s period-2 indirect utility given that $O_1$ is the government in period 2 (i.e. $I_2=O_1$) is
\begin{equation}
\label{IUI}
W^{O_1}(\tau_2|I_2=O_1) =(1-\tau_2)m +  \vfrac{2\tau_2m}{(1+\sigma^D)}
\end{equation}
When $I_1$ remains in power in period 2 (i.e. $I_2=I_1$), $O_1$'s period-2 indirect utility is given by
\begin{equation}
\label{IUO}
W^{O_1}(\tau_2|I_2=I_1) =(1-\tau_2)m + \vfrac{2\sigma^D\tau_2m}{(1+\sigma^D)}
\end{equation}
 For $I_2=F$, substituting $ r_2^F=\vfrac{2\tau_2 m}{(2+\sigma^F)}$ into $r_2^{O_1}=\sigma^Fr_2^F$, and using the result in (\ref{us}), we have that $O_1$'s period-2 indirect utility is 
\begin{equation}
\label{IUOF}
W^{O_1}(\tau_2|I_2=F)=(1-\tau_2)m+\vfrac{2\sigma^F\tau_2 m}{(2+\sigma^F)}
\end{equation}

Now I compute  $O_1$'s expected utility. Combining (\ref{IUI}), (\ref{IUO}) and (\ref{IUOF}), with $\alpha$ representing the probability of interstate conflict, we have that $O_1$'s expected utility (in period 2) in the event of a civil war is
\small
\begin{equation}
\label{UBID}
\begin{split}
\alpha\Big[(\omega+\rho\lambda)  W^{O_1}(\tau_2|I_2=O_1)+(1-\omega-\rho)W^{O_1}(\tau_2|I_2=I_1)+\rho(1-\lambda)  W^{O_1}(\tau_2|I_2=F)\Big] \\+(1-\alpha)\Big[\delta  W^{O_1}(\tau_2|I_2=O_1)+(1-\delta) W^{O_1}(\tau_2|I_2=I_1)\Big]
\end{split}
\end{equation}
and in the event of internal peace is
\begin{equation}
\label{UBNID}
\begin{split}
\alpha\Big[\mu\lambda W^{O_1}(\tau_2|I_2=O_1)+(1-\mu)W^{O_1}(\tau_2|I_2=I_1)+  \mu(1- \lambda) W^{O_1}(\tau_2|I_2=F)\Big]\\+(1-\alpha)\Big[\epsilon  W^{O_1}(\tau_2|I_2=O_1)+(1-\epsilon)W^{O_1}(\tau_2|I_2=I_1)\Big]
\end{split}
\end{equation}
Note that $O_1$ decides whether or not to trigger a civil war by comparing (\ref{UBID}) and (\ref{UBNID}). Rearranging these expressions (and combining them with (\ref{IUI}), (\ref{IUO}) and (\ref{IUOF})), it is easy to see that $O_1$ triggers a civil war when $\sigma^F>\overline{\sigma}^F$. 
\end{proof}


\begin{proof}[\textbf{Proof that a civil war is more likely for low values of $\sigma^D$}]  To see this, we can differentiate $\overline{\sigma}^F$ in (\ref{sigmaFthreshold}) with respect to $\sigma^D$, and rearrange to get 
 \vspace{-0.1cm}
\small
 \begin{equation}
\label{ }
\frac{2\alpha(\rho-\mu)(1-\lambda) (\alpha(\rho-\mu)(1+\lambda) +2(\alpha\omega+(1-\alpha)\delta-(1-\alpha)\epsilon))}{(\alpha(\rho-\mu)(1 -\sigma^D\lambda) +(1-\sigma^D)(\alpha\omega+(1-\alpha)\delta-(1-\alpha)\epsilon ))^2}>0
\end{equation}
\vspace{-0.2cm}
where we have used that $\rho-\mu>0$ (Eq. (\ref{phomu})). 

\end{proof}


\begin{proof}[\textbf{Proof that an increase in $\lambda$ increases the chance of a civil war}]   To see this, we can differentiate $\overline{\sigma}^F$ in (\ref{sigmaFthreshold}) with respect to $\lambda$, and rearrange to get
\small
\begin{equation}
\label{dfproblamda}
-\frac{2\alpha(\rho-\mu)(1-(\sigma^D)^2)(\alpha(\rho-\mu) +\alpha\omega+(1-\alpha)\delta-(1-\alpha)\epsilon)}{(\alpha(\rho-\mu)(1 -\sigma^D\lambda) +(1-\sigma^D)(\alpha\omega+(1-\alpha)\delta-(1-\alpha)\epsilon ))^2}<0
\end{equation}
where we have used that $\rho-\mu>0$ (Eq. (\ref{phomu})). 

\end{proof}


\begin{proof}[\textbf{Proof that an increase in $\alpha$ may increase or decrease the chance of a civil war}]  To see this, we can differentiate $\overline{\sigma}^F$ in (\ref{sigmaFthreshold}) with respect to $\alpha$, and rearrange to get
\small
\begin{equation}
\label{partiagammaalpha}
\frac{2(\rho-\mu) (1-\lambda)(\delta -\epsilon )(1-\sigma^D)(1+\sigma^D)}{(\alpha(\rho-\mu)(1 -\sigma^D\lambda)  +(1-\sigma^D)(\alpha\omega+(1-\alpha)\delta-(1-\alpha)\epsilon ))^2} \gtrless 0.
\end{equation}
Note that this expression is positive when $\delta >\epsilon $, and is negative when $\epsilon >\delta $ (note that $\rho-\mu>0$ by  (\ref{phomu})). 

\end{proof}


\begin{proof}[\textbf{Proof of Proposition 1}] I look at $\frac{\partial \phi}{\partial \alpha}$, where $\phi$ is given by (\ref{probturnov}).  When $\sigma^F\leq \overline{\sigma}^F$, $\gamma=0$, so $\phi=\alpha\mu+(1-\alpha)\epsilon$.  Differentiating this expression with respect to $\alpha$ gives us
\begin{equation}
\label{ }
\frac{\partial \phi}{\partial \alpha} =-(\epsilon-\mu)
\end{equation}
 which by (\ref{epsilonmu}) is always negative. Thus, when  $\sigma^F\leq\overline{\sigma}^F$, an increased risk of external conflict decreases $\phi$. 

\medskip

When $\sigma^F> \overline{\sigma}^F$, $\gamma=1$, so $\phi=\alpha\omega+(1-\alpha)\delta+\alpha\rho$.  Differentiating this expression with respect to $\alpha$ gives us
\begin{equation}
\label{ }
\frac{\partial \phi}{\partial \alpha}=\omega -\delta +\rho >0
\end{equation}
where we have used that $\omega -\delta>0$  (Eq. (\ref{phiowwp})). Thus, when  $\sigma^F>\overline{\sigma}^F$, an increased risk of external conflict increases $\phi$. 

\end{proof}


\begin{proof}[\textbf{Proof of Proposition 2}] 

First, I compute period-2 optimal policies. They are similar to the period-1 optimal policies; the only difference is that the budget constraint in period 1 includes the costs associated with investment. $I_1$ chooses $\{t_1, \textbf{r}_1=\{r_1^A,r_1^B,r_1^F\}\}$ to maximize (\ref{us}), subject to (\ref{BC}) and $t_1\leq \tau_1$. It is easy to see that we have a corner solution, with $t_1=\tau_1$, $r_1^F=0$, $r_1^{O_1}= \sigma^Dr_1^{I_1}$ and 
 $r_1^{I_1}=\vfrac{2(\tau_1m-C(\tau_2-\tau_1))}{(1+\sigma^D)}$. 

\bigskip

I now study the decision to invest in fiscal capacity. First, I compute $I_1$'s first- and second-period indirect utilities. When $I_1$ remains in power in period 2 (i.e. $I_2=I_1$), $I_1$'s period-2 utility is
\begin{equation}
\label{IUII}
W^{I_1}(\tau_2|I_2=I_1) =(1-\tau_2)m +  \vfrac{2\tau_2m}{(1+\sigma^D)}
\end{equation}
When $O_1$ gains power in period 2 (i.e. $I_2=O_1$), $I_1$'s period-2 utility  is
\begin{equation}
\label{IUOI}
W^{I_1}(\tau_2|I_2=O_1) =(1-\tau_2)m +  \vfrac{2\sigma^D\tau_2m}{(1+\sigma^D)}
\end{equation}
And  when $F$ governs in period 2 (i.e. $I_2=F$), $I_1$'s period-2 indirect utility is 
\begin{equation}
\label{IUOFI}
W^{I_1}(\tau_2|I_2=F)=(1-\tau_2)m
\end{equation}
And $I_1$'s first-period indirect utility is
\begin{equation}
\label{IUFP}
W^{I_1}(\tau_1,C(\tau_2-\tau_1))=(1-\tau_1)m +  \vfrac{2(\tau_1m-C(\tau_2-\tau_1))}{(1+\sigma^D)}
\end{equation}

\medskip
I now calculate $I_1$'s expected utility. Combining (\ref{IUII}), (\ref{IUOI}), (\ref{IUOFI}) and (\ref{IUFP}), we have that $I_1$'s expected utility when there is no civil war is 
\small
\begin{equation}
\label{I1expnocw}
\begin{split}
W^{I_1}(\tau_1,C(\tau_2-\tau_1))\\+\alpha\Big[(1-\mu)W^{I_1}(\tau_2|I_2=I_1)+\mu\lambda W^{I_1}(\tau_2|I_2=O_1)+  \mu(1- \lambda) W^{I_1}(\tau_2|I_2=F)\Big]\\
+(1-\alpha)\Big[(1-\epsilon)W^{I_1}(\tau_2|I_2=I_1)+\epsilon W^{I_1}(\tau_2|I_2=O_1)\Big]
\end{split}
\end{equation}
When there is a civil war, $I_1$'s expected utility is
\small
\begin{equation}
\label{I1expyescw}
\begin{split}
W^{I_1}(\tau_1,C(\tau_2-\tau_1))\\+\alpha\Big[(1-\omega-\rho)W^{I_1}(\tau_2|I_2=I_1)+(\omega+\rho\lambda) W^{I_1}(\tau_2|I_2=O_1)+\rho(1-\lambda) W^{I_1}(\tau_2|I_2=F)\Big]\\
+(1-\alpha)\Big[(1-\delta)W^{I_1}(\tau_2|I_2=I_1)+\delta W^{I_1}(\tau_2|I_2=O_1)\Big]
\end{split}
\end{equation}
When $I_1$ expects that there won't be a civil war (i.e. $\sigma^F\leq \overline{\sigma}^F$),  $I_1$ chooses $\tau_2$ at the beginning of period 1 by maximizing (\ref{I1expnocw}). Differentiating (\ref{I1expnocw}) with respect to $\tau_2$, and setting it equal to zero, gives us the first-order condition:
\small
 \begin{equation}
 \begin{split}
 \label{FOCpeace}
-W_C^{I_1}(\tau_1,C(\tau^*_2-\tau_1))C_\tau(\tau^*_2-\tau_1)  \geq \\
+\alpha \Big[(1-\mu)W_\tau^{I_1}(\tau^*_22|I_2=I_1)+\mu\lambda W_\tau^{I_1}(\tau^*_2|I_2=O_1)+  \mu(1- \lambda) W_\tau^{I_1}(\tau^*_2|I_2=F)\Big]\\
+(1-\alpha)\Big[(1-\epsilon)W_\tau^{I_1}(\tau^*_2|I_2=I_1)+\epsilon W_\tau^{I_1}(\tau^*_2|I_2=O_1)\Big]
\end{split}
\end{equation}
which, rearranging, and combining it with (\ref{IUII}), (\ref{IUOI}), (\ref{IUOFI}) and (\ref{IUFP}), is equivalent to  
 \begin{equation}
 \begin{split}
 \label{FOCpeace0}
2C_\tau(\tau^*_2-\tau_1)  \geq 2m\big(1-(1-\alpha)\epsilon(1-\sigma^D)-\alpha \mu(1-\lambda\sigma^D)\big)-m(1+\sigma^D).
\end{split}
\end{equation}
From  (\ref{probturnov}), we have that for no civil war,  $\phi=\alpha\mu+(1-\alpha)\epsilon$. Thus, (\ref{FOCpeace0}) is equivalent to
 \begin{equation}
 \begin{split}
 \label{FOCpeace1}
2C_\tau(\tau^*_2-\tau_1)  \geq 2m\big(1-\phi(1-\sigma^D)-\alpha \mu(1-\lambda)\sigma^D\big)-m(1+\sigma^D).
\end{split}
\end{equation}
When $I_1$ expects there will be a civil war (i.e. $\sigma^F>\overline{\sigma}^F$),  $I_1$  chooses $\tau_2$ at the beginning of period 1 by maximizing (\ref{I1expyescw}). Differentiating (\ref{I1expyescw}) with respect to $\tau_2$, and setting it equal to zero, gives us the first-order condition:
 \begin{equation}
 \begin{split}
-W_C^{I_1}(\tau_1,C(\tau^*_2-\tau_1))C_\tau(\tau^*_2-\tau_1)  \geq \\
\alpha\Big[(1-\omega-\rho)W_\tau^{I_1}(\tau^*_2|I_2=I_1)+(\omega+\rho\lambda)  W_\tau^{I_1}(\tau^*_2|I_2=O_1)+\rho(1-\lambda) W_\tau^{I_1}(\tau^*_2|I_2=F)\Big]\\ 
 +(1-\alpha)\Big[(1-\delta)W_\tau^{I_1}(\tau^*_2|I_2=I_1)+\delta W_\tau^{I_1}(\tau^*_2|I_2=O_1)\Big]
\end{split}
\end{equation}
which, rearranging, and combining it with (\ref{IUII}), (\ref{IUOI}), (\ref{IUOFI}) and (\ref{IUFP}), is equivalent to  
 \begin{equation}
 \begin{split}
2C_\tau(\tau^*_2-\tau_1)\geq 2m\Big[1-(\alpha\omega+(1-\alpha)\delta+\alpha\rho)+\sigma^D(\alpha\omega+(1-\alpha)\delta+\alpha\rho\lambda)\Big]-m(1+\sigma^D).
\end{split}
\end{equation}
From  (\ref{probturnov}), we have that for civil war,  $\phi=\alpha\omega+(1-\alpha)\delta+\alpha\rho$, thus the last expression is equal to
 \begin{equation}
\label{FOCwar1}
 \begin{split}
2C_\tau(\tau^*_2-\tau_1)\geq 2m\Big[1-\phi(1-\sigma^D)-\alpha\rho(1-\lambda)\sigma^D\Big]-m(1+\sigma^D).
\end{split}
\end{equation}
Combining the expressions in (\ref{FOCpeace1}) and (\ref{FOCwar1}), recalling that $\gamma=1$ if $\sigma^F>\overline{\sigma}^F$ and $\gamma=0$ otherwise, we have that 
 \begin{equation}
\label{FOCpeacewarapp}
 \begin{split}
2C_\tau(\tau^*_2-\tau_1)\geq 2m\Big[1-\phi(1-\sigma^D)-\alpha(\gamma\rho+(1-\gamma) \mu)(1-\lambda)\sigma^D\Big]-m(1+\sigma^D).
\end{split}
\end{equation}
Note that for $\tau^*_2>\tau_1$, the complementary slackness condition implies that (\ref{FOCpeacewarapp}) holds with equality. Solving for $\tau^*_2$, 
we get (\ref{tau2fin}).  Note that differentiating (\ref{FOCpeace1}) or (\ref{FOCwar1}) again with respect to $\tau_2$ gives us the second-order condition $-2C_{\tau\tau}(\tau_2-\tau_1)<0$, which is always true, given that $C(\cdot)$ is strictly convex; thus,  (\ref{tau2fin}) gives a maximum.

From (\ref{FOCpeacewarapp}), note that $\tau^*_2$ depends on whether $\gamma=1$ or $\gamma=0$. In equilibrium, when $\sigma^F\leq\overline{\sigma}^F$ (i.e., when $\gamma=0$),  $O_1$ decides not to trigger a civil war, and $I_1$ chooses $\tau^*_2$ according to  (\ref{tau2fin}) for $\gamma=0$.  When $\sigma^F>\overline{\sigma}^F$  (i.e., when $\gamma=1$),  $O_1$ decides to trigger a civil war, and $I_1$  chooses $\tau^*_2$ according to (\ref{tau2fin}) for $\gamma=1$.

\medskip
Now I look at $\frac{\partial \tau^*_2}{\partial \alpha}$. For the case of $\sigma^F\leq \overline{\sigma}^F$, since $\gamma=0$, we have that (\ref{tau2fin}) is equal to
\begin{equation}
\label{tau2fingamma0}
\tau^*_2 = C_\tau^{-1}\big(m\big[-(\alpha\mu+(1-\alpha)\epsilon)(1-\sigma^D)-\alpha \mu(1-\lambda)\sigma^D+\vfrac{(1-\sigma^D)}{2}\big]\big) +\tau_1.
\end{equation}
Differentiating (\ref{tau2fingamma0}) with respect to $\alpha$,  it is easy to see that, given the strict convexity of $C(\cdot)$, and that $m$, $\sigma^D$ and $\tau_1$ are constants, the sign of $\frac{\partial \tau^*_2}{\partial \alpha}$ depends on the sign of 
\begin{equation}
\label{ }
\frac{\partial }{\partial \alpha}(-(\alpha\mu+(1-\alpha)\epsilon)(1-\sigma^D)-\alpha \mu(1-\lambda)\sigma^D)=(\epsilon-\mu)(1-\sigma^D)- \mu(1-\lambda)\sigma^D
\end{equation}
which, rearranging,  is greater than zero if and only if 
\begin{equation}
\label{deltataua1}
(\epsilon-\mu)>\sigma^D(\epsilon-\lambda\mu).
\end{equation}
Thus, when  $\sigma^F\leq\overline{\sigma}^F$, depending on whether $(\epsilon-\mu)\gtrless\sigma^D(\epsilon-\lambda\mu)$, an increase in the risk of external conflict can increase,  decrease or not affect investment in fiscal capacity. 

\medskip

For the case of $\sigma^F>\overline{\sigma}^F$, since $\gamma=1$, we have that (\ref{tau2fin}) is equal to
\begin{equation}
\label{tau2fingamma1}
\tau^*_2 = C_\tau^{-1}\big(m\big[-(\alpha\omega+(1-\alpha)\delta+\alpha\rho)(1-\sigma^D)-\alpha\rho(1-\lambda)\sigma^D+\vfrac{(1-\sigma^D)}{2}\big]\big) +\tau_1.
\end{equation}
Differentiating (\ref{tau2fingamma1}) with respect to $\alpha$, we have that the sign of $\frac{\partial \tau^*_2}{\partial \alpha}$ now depends on the sign of 
\begin{equation}
\label{deltataua1}
\frac{\partial }{\partial \alpha}(-(\alpha\omega+(1-\alpha)\delta+\alpha\rho)(1-\sigma^D)-\alpha\rho(1-\lambda)\sigma^D)=-(\omega-\delta+\rho)(1-\sigma^D)-\rho(1-\lambda)\sigma^D
\end{equation}
which by (\ref{phiowwp}) is always negative. Thus, when  $\sigma^F>\overline{\sigma}^F$, an increased risk of an external conflict always decreases investments in fiscal capacity. 

\end{proof}


\begin{proof}[\textbf{Proof of Proposition 3}] For the case of $\sigma^F> \overline{\sigma}^F$, it is clear from Propositions 1.A and 2.A together that an increased risk of external conflict  always implies both increased political turnover and decreased investment in fiscal capacity. For the case of $\sigma^F\leq \overline{\sigma}^F$, Propositions 1.A.2, 1.A.3 and 2.B together imply that an increased risk of external conflict, even though it always implies a decrease in political turnover, may not affect (or may even decrease) investment in fiscal capacity. This happens when, in addition to $\sigma^F\leq \overline{\sigma}^F$, we have that $(\epsilon-\mu)\leq \sigma^D(\epsilon-\lambda\mu)$. 
\end{proof}


\begin{proof}[\textbf{Proof of Proposition 4}] 

First, we consider  $O_1$'s choice given an offer $\sigma_2^{D^*}$ by $I_1$. If $O_1$ rejects the offer, $O_1$ expects $r_2^{O_2}=0$ for $I_2\in\{A,B\}$, since $\sigma_2^D=0$. However, this does not mean that $r_2^{O_1}=0$ for all cases, since it could be that $I_2=F$, with which $O_1$  could receive positive transfers from $F$ in period 2 or be in power thanks to $F$. From (\ref{sigmaFthreshold}), we know that if $\sigma_2^D=0$, $O_1$ will start a civil war. In this scenario, the reservation utility of $O_1$ will be 
\begin{equation}
\label{ }
\begin{split}
(\alpha\omega+\alpha\rho\lambda+(1-\alpha)\delta) W^{O_1}(\tau_2|I_2=O_1)+(\alpha-\alpha\omega-\alpha\rho+(1-\alpha)(1-\delta))W^{O_1}(\tau_2|I_2=I_1)\\+ \alpha\rho(1-\lambda) W^{O_1}(\tau_2|I_2=F)
\end{split}
\end{equation}
which, from (\ref{IUI}), (\ref{IUO}) and (\ref{IUOF}), is equivalent to 
\begin{equation}
\label{prop5outside1}
2\tau_2m(\alpha\omega+\alpha\rho\lambda+(1-\alpha)\delta+\alpha\rho(1-\lambda)  \vfrac{\sigma^F}{(2+\sigma^F)})+(1-\tau_2)m.
\end{equation}
If $O_1$ accepts $I_1$'s offer, $O_1$ will not trigger a civil war, with which it  expects to get
\begin{equation}
\label{ }
\begin{split}
(\alpha\mu\lambda'+(1-\alpha)\epsilon)  W^{O_1}(\tau_2|I_2=O_1)+(\alpha(1-\mu)+(1-\alpha)(1-\epsilon)) W^{O_1}(\tau_2|I_2=I_1) \\+\alpha\mu(1- \lambda') W^{O_1}(\tau_2|I_2=F)
\end{split}
\end{equation}
which, again from (\ref{IUI}), (\ref{IUO}) and (\ref{IUOF}), is equivalent to
\begin{equation}
\label{prop5inside1}
2\tau_2m\Big([(\alpha\mu\lambda+(1-\alpha)\epsilon)+(\alpha(1-\mu)+(1-\alpha)(1-\epsilon))\sigma_2^{D^*}]/(1+\sigma_2^{D^*})+\alpha\mu(1- \lambda)\sigma^F/(2+\sigma^F)\Big)+(1-\tau_2)m.
\end{equation}
Thus, $O_1$ accepts the offer if the expression in (\ref{prop5inside1}) is greater than or equal to the expression in (\ref{prop5outside1}); this condition is equivalent to
\begin{equation}
\label{prop5condacc1}
\begin{split}
\alpha(\omega+(\rho-\mu)\lambda)+\alpha(\rho-\mu)(1- \lambda)\sigma^F/(2+\sigma^F)\leq \\\sigma_2^{D^*}(1-\alpha(\omega+\mu+\rho\lambda)-(1-\alpha)2\epsilon-\alpha(\rho-\mu)(1- \lambda)\sigma^F/(2+\sigma^F))
\end{split}
\end{equation}
where we have used the assumption that $\epsilon=\delta$. Now we study the conditions under which it is optimal for $I_1$ to offer $\sigma_2^{D^*}>0$ that  $O_1$ will accept. Note that if $I_1$ offers $\sigma_2^{D}$ such that (\ref{prop5condacc1}) is satisfied, in the second period $I_1$ will get
\begin{equation}
\label{objc14a}
\begin{split}
\alpha\Big[(1-\mu)W^{I_1}(\tau_2|I_2=I_1)+\mu\lambda W^{I_1}(\tau_2|I_2=O_1)+  \mu(1- \lambda) W^{I_1}(\tau_2|I_2=F)\Big]\\
+(1-\alpha)\Big[(1-\epsilon)W^{I_1}(\tau_2|I_2=I_1)+\epsilon W^{I_1}(\tau_2|I_2=O_1)\Big].
\end{split}
\end{equation}
 Replacing  (\ref{IUII}), (\ref{IUOI}) and (\ref{IUOFI}) in (\ref{objc14a}), and rearranging, we have that (\ref{objc14a}) is equal to 
\begin{equation}
\label{objc14a1}
\begin{split}
(1-\tau_2)m+2m\tau_2\big[1-\alpha\mu(1-\lambda \sigma_2^D)-(1-\alpha)\epsilon(1-\sigma_2^D)\big]/(1+\sigma_2^D).
\end{split}
\end{equation}
For the case in which $I_1$ chooses a $\sigma_2^{D}$ such that (\ref{prop5condacc1}) is satisfied, we have several possibilities, depending on the signs of the terms on the left and right sides of (\ref{prop5condacc1}) (and on whether $(1-\alpha)\epsilon \gtrless  \vfrac{1}{2}$). First, note that differentiating (\ref{objc14a1}) with respect to $\sigma_2^{D}$, we obtain the first-order condition
\begin{equation}
\label{focappprop4}
2\tau_2 m(2(1-\alpha)\epsilon-1+\alpha\mu(1+\lambda))/(1+\sigma_2^{D})^2\leq 0
\end{equation}
with the complementary slackness conditions associated with (\ref{prop5condacc1}), $\sigma_2^{D}\geq 0$ and $\sigma_2^{D}\leq 1$.  Note that the expression in (\ref{focappprop4}) is strictly negative when $2(1-\alpha)\epsilon-1+\alpha\mu(1+\lambda)<0$, or, equivalently, when $1/2>(1-\alpha)\epsilon+\alpha\mu(1+\lambda)/2$. We assume that this last condition is satisfied: $1/2>(1-\alpha)\epsilon$ is plausible since each group makes up half of the population, and with respect to the additional term, we basically assume that $F$ and $O_1$ cannot together be extremely powerful. 

If the term on the left side of (\ref{prop5condacc1}) is strictly positive (i.e. $\alpha(\omega+(\rho-\mu)\lambda)+\alpha(\rho-\mu)(1- \lambda)\sigma^F/(2+\sigma^F)>0$), then (\ref{prop5condacc1}) implies that $1-\alpha(\omega+\mu+\rho\lambda)-(1-\alpha)2\epsilon-\alpha(\rho-\mu)(1- \lambda)\sigma^F/(2+\sigma^F)$ should be also strictly positive. 
If we look at a solution in which $\sigma_2^{D}> 0$, these conditions, as well as (\ref{focappprop4}) and the complementary slackness conditions imply that (\ref{prop5condacc1})  is binding, so we have  (\ref{condprop4a}), i.e. that 
\begin{equation}
\label{sigma2barg}
\sigma_2^{D^*}=\frac{\alpha(\omega+(\rho-\mu)\lambda)+\alpha(\rho-\mu)(1- \lambda)\sigma^F/(2+\sigma^F)}{1-\alpha(\omega+\mu+\rho\lambda)-2(1-\alpha)\epsilon-\alpha(\rho-\mu)(1- \lambda)\sigma^F/(2+\sigma^F)}.
\end{equation}
To identify the cases in which $\sigma_2^{D^*}<1$ and $\sigma_2^{D^*}=1$, note that by (\ref{sigma2barg}) $\sigma_2^{D^*}<1$  occurs when 
\begin{equation}
\label{prop4f1stconda}
\alpha\omega+\alpha\rho\lambda+\alpha\mu(1-\lambda)/2+(1-\alpha)\epsilon+\alpha(\rho-\mu)(1- \lambda)\sigma^F/(2+\sigma^F)<1/2.
\end{equation}
In this case, note also that (\ref{prop4f1stconda}) implies that $1-\alpha(\omega+\mu+\rho\lambda)-2(1-\alpha)\epsilon-\alpha(\rho-\mu)(1- \lambda)\sigma^F/(2+\sigma^F)>0$ (under the assumption that  $1/2>(1-\alpha)\epsilon+\alpha\mu(1+\lambda)/2$), so the first condition is sufficient. As for $\sigma_2^{D^*}=1$, note from (\ref{sigma2barg})  that we must have $\alpha\omega+\alpha\rho\lambda+\alpha\mu(1-\lambda)/2+(1-\alpha)\epsilon+\alpha(\rho-\mu)(1- \lambda)\sigma^F/(2+\sigma^F)=\vfrac{1}{2}$. This last equality, combined with (\ref{prop4f1stconda}), provides the conditions under which $0\leq \sigma_2^{D^*}\leq 1$, where $\sigma_2^{D^*}$ is given by (\ref{sigma2barg}). 

As for $\sigma_2^{D^*}=0$, note that this occurs when (\ref{prop5condacc1})  is not binding. Note that this can only be true if $1-\alpha(\omega+\mu+\rho\lambda)-2(1-\alpha)\epsilon-\alpha(\rho-\mu)(1- \lambda)\sigma^F/(2+\sigma^F)>0$ and $\alpha(\omega+(\rho-\mu)\lambda)+\alpha(\rho-\mu)(1- \lambda)\sigma^F/(2+\sigma^F)\leq 0$ (note that it cannot be true that both are negative or positive and (\ref{prop5condacc1})  is not binding). In this case, since as we previously showed, (\ref{prop4f1stconda}) and $1/2>(1-\alpha)\epsilon+\alpha\mu(1+\lambda)/2$ imply that $1-\alpha(\omega+\mu+\rho\lambda)-2(1-\alpha)\epsilon-\alpha(\rho-\mu)(1- \lambda)\sigma^F/(2+\sigma^F)>0$, to have $\sigma_2^{D^*}>0$ in addition to (\ref{prop4f1stconda}), we need 
\begin{equation}
\label{prop4f1stcondb}
(1-\alpha)\epsilon+\alpha\mu(1+\lambda)/2<\alpha\omega+\alpha\rho\lambda+\alpha\mu(1-\lambda)/2+(1-\alpha)\epsilon+\alpha(\rho-\mu)(1- \lambda)\sigma^F/(2+\sigma^F).
\end{equation}
Finally, note that when (\ref{prop4f1stconda}) does not hold, (\ref{prop5condacc1}) is not satisfied. 

\medskip

Now we examine under what additional conditions $I_1$ will offer $\sigma_2^{D^*}>0$, and when it will offer $\sigma_2^{D^*}=0$ (or something that $O_1$ will reject). Note that if $O_1$ rejects the offer,  $I_1$ will get 
\footnotesize
\begin{equation}
\label{objc14b}
\begin{split}
\alpha\Big[(1-\omega-\rho)W^{I_1}(\tau_2|I_2=I_1,\sigma_2^{D}=0)+(\omega+\rho\lambda) W^{I_1}(\tau_2|I_2=O_1,\sigma_2^{D}=0)+\rho(1-\lambda)  W^{I_1}(\tau_2|I_2=F)\Big]\\+(1-\alpha)\Big[(1-\delta)W^{I_1}(\tau_2|I_2=I_1,\sigma_2^{D}=0)+\delta W^{I_1}(\tau_2|I_2=O_1,\sigma_2^{D}=0)\Big]
\end{split}
\end{equation}
\small
where the functions $W^{I_1}(\cdot)$ are still defined by  (\ref{IUII}), (\ref{IUOI}) and (\ref{IUOFI}), but now  $\sigma_2^D=0$. 
Replacing (\ref{IUII}), (\ref{IUOI}) and (\ref{IUOFI}), $\delta=\epsilon$, and rearranging, we have that (\ref{objc14b}) is equal to
\begin{equation}
\label{objc14a1b}
(1-\tau_2) m+2\tau_2 m\alpha(1-\omega-\rho )+2\tau_2 m(1-\alpha)(1-\epsilon).
\end{equation}
Thus $I_1$ prefers to offer $\sigma_2^{D^*}>0$ if the expression in (\ref{objc14a1}) calculated at $\sigma_2^{D^*}$ is greater than the expression in (\ref{objc14a1b}), which, rearranging, is equivalent to
\begin{equation}
\label{objc14c}
\alpha(\omega+\rho-\mu)>\sigma_2^{D^*}(1-\alpha(\omega+\rho+\mu\lambda)-2(1-\alpha)\epsilon).
\end{equation}
For the case where $\sigma_2^{D^*}\in(0,1)$, note that (\ref{prop4f1stcondb}) implies that the left side in (\ref{objc14c}) is always greater than zero. Replacing $\sigma_2^{D^*}$ from (\ref{sigma2barg}) in (\ref{objc14c}), and rearranging, we have that (\ref{objc14c}) becomes
\begin{equation}
\label{ }
1-\alpha\mu(1+\lambda)-2(1-\alpha)\epsilon>0
\end{equation}
which is equivalent to $1/2>(1-\alpha)\epsilon+\alpha\mu(1+\lambda)/2$, which we assumed to be true.  Thus, for this case we can conclude that if  (\ref{prop4f1stconda}) and (\ref{prop4f1stcondb}) are satisfied ((\ref{condprop40})  and (\ref{condprop4}) in the main text), there is a unique $\sigma_2^{D^*}\in(0,1)$ given by (\ref{sigma2barg}) ((\ref{condprop4a}) in the main text) that will be proposed by  $I_1$ and accepted by $O_1$. 

For the case of $\sigma_2^{D^*}=1$, note that (\ref{objc14c}) is equivalent to  
\begin{equation}
\label{condsig1prop4}
\alpha(\omega+\rho)-\alpha\mu(1-\lambda)/2+(1-\alpha)\epsilon> \vfrac{1}{2}
\end{equation}
which holds when (\ref{condprop40}) is satisfied with equality. 
\end{proof} 


\begin{proof}[\textbf{Proof of Proposition 5}]  First, note that since in this scenario it may that $\sigma_2^D\neq \sigma_1^D$, we need to generalize the expression for  $\tau_2^*$ in (\ref{tau2fin}) to account for this possibility. Repeating the procedure through which we got (\ref{tau2fin}) (see the proof of Prop. 2) but distinguishing between $\sigma_1^D$ and  $\sigma_2^D$,  we get 
 \small
\begin{equation}
\label{tau2fin0}
\tau^*_2 = C_\tau^{-1}\Bigg(m\frac{(1+\sigma_1^D)}{(1+\sigma_2^D)}\Big[-\phi(1-\sigma^D)-\alpha(\gamma\rho+(1-\gamma) \mu)(1-\lambda)\sigma^D+\vfrac{(1-\sigma_2^D)}{2}\Big]\Bigg) +\tau_1.
\end{equation}
Now we examine the effect of an increase in the risk of external conflict on $\tau^*_2$. 
First, we examine the case in which (\ref{condprop40}) and (\ref{condprop4}) hold and $\sigma_2^{D*}<1$. From the proof of Prop. 4, note that $\sigma_2^{D*}<1$ when the inequality in (\ref{condprop40}) is strict.  In this scenario, there is no civil war, so $\gamma=0$, and $\sigma_2^D$ is given by (\ref{sigma2barg}) (or (\ref{condprop4a}) in the main text). Also note that by assumption, $\sigma_1^D=0$.  Differentiating (\ref{tau2fin0}) with respect to $\alpha$, we have that the sign of $\vfrac{\partial \tau^*_2}{\partial \alpha}$ depends on the sign of 
\begin{equation}
\label{deltatauappprop5b}
\frac{1}{(1+\sigma_2^{D^*})^2}\Big((1+\sigma_2^{D^*})(\epsilon(1-\sigma_2^{D^*})- \mu(1-\lambda\sigma_2^{D^*}))-(1-2(1-\alpha)\epsilon -\alpha \mu(1+\lambda))\frac{\partial \sigma_2^{D^*}}{\partial \alpha}\Big)
\end{equation}
where $\sigma_2^{D^*}$ is given by (\ref{condprop4a}). From (\ref{deltatauappprop5b}), note that  $\vfrac{\partial \tau^*_2}{\partial \alpha}> 0$ if and only if  
\begin{equation}
\label{condfprop5}
(1+\sigma_2^{D^*})(\epsilon(1-\sigma_2^{D^*})- \mu(1-\lambda\sigma_2^{D^*}))> (1-2(1-\alpha)\epsilon -\alpha \mu(1+\lambda))\frac{\partial \sigma_2^{D^*}}{\partial \alpha}.
\end{equation}
From this condition, note that the term on the right is non-negative. To see this, first note that by assumptions $\vfrac{1}{2}\geq\epsilon$ and (\ref{epsilonmu}), $1-2(1-\alpha)\epsilon -\alpha \mu(1+\lambda)>0$. As for $\vfrac{\partial \sigma_2^{D^*}}{\partial \alpha}$, differentiating (\ref{condprop4a}) with respect to $\alpha$ we get
\begin{equation}
\label{partialsigmahat}
 \frac{\partial \sigma^{D^*}_2}{\partial \alpha}=\frac{(1-2\epsilon)(\omega+\lambda(\rho-\mu)+(\rho-\mu)(1- \lambda)\sigma^F/(2+\sigma^F))}{(1-\alpha(\omega+\mu+\rho\lambda)-2(1-\alpha)\epsilon-\alpha(\rho-\mu)(1- \lambda)\sigma^F/(2+\sigma^F))^2}
\end{equation}
and note that the numerator is non-negative because of assumption (\ref{phomu}) and because $\vfrac{1}{2}\geq\epsilon$. 

As for the term on the left in (\ref{condfprop5}), note that its sign depends on whether $\epsilon(1-\sigma_2^{D^*})\gtrless \mu(1-\lambda\sigma_2^{D^*})$, or equivalently, on whether $(\epsilon-\mu)\gtrless\sigma_2^{D^*}(\epsilon-\lambda\mu)$. Also note that when $\epsilon(1-\sigma_2^{D^*})<\mu(1-\lambda\sigma_2^{D^*})$, since the term on the right in \ref{condfprop5}) is  non-negative, $\vfrac{\partial \tau^*_2}{\partial \alpha}<0$, i.e. an increased risk of external conflict decreases investment in fiscal capacity. 

To examine the case where $\epsilon(1-\sigma_2^{D^*})>\mu(1-\lambda\sigma_2^{D^*})$,  we replace (\ref{partialsigmahat}) in (\ref{deltatauappprop5b}), and rearrange, with which we have that (\ref{deltatauappprop5b}) is equivalent to
\scriptsize
\begin{equation}
\label{ }
-\Big(1-\alpha\mu(1+\lambda)-2(1-\alpha)\epsilon\Big)\frac{\big(1-\alpha\mu(1+\lambda)-2(1-\alpha)\epsilon\big)\big((\omega-\epsilon)+\mu+(\rho-\mu)(\lambda+(1- \lambda)\sigma^F/(2+\sigma^F))\big)+2\alpha\lambda\mu(\epsilon-\mu\lambda)}{(1-\alpha(\omega+\mu+\rho\lambda)-2(1-\alpha)\epsilon-\alpha(\rho-\mu)(1- \lambda)\sigma^F/(2+\sigma^F))^2}
\end{equation}
\small
which is less than zero since: (i) $1-\alpha\mu(1+\lambda)-2(1-\alpha)\epsilon>0$ (implied by $\vfrac{1}{2}\geq\epsilon$ and (\ref{epsilonmu})), (ii) $\rho>\mu$ (by (\ref{phomu})),  (iii) $\omega>\epsilon=\delta$ (by (\ref{phiowwp})), and (iv) $\epsilon>\mu$ (because it is implied by $\epsilon(1-\sigma_2^{D^*})>\mu(1-\lambda\sigma_2^{D^*})$).  This shows that  when  (\ref{condprop40}) and (\ref{condprop4})  hold, and the inequality in (\ref{condprop40}) is strict, $\vfrac{\partial \tau^*_2}{\partial \alpha}<0$, so an increased risk of external conflict does not increase investments in fiscal capacity. 

Note that from the proof of Prop. 4, $\sigma_2^{D^*}=1$ occurs when (\ref{condprop40}) and (\ref{condprop4})  hold, and (\ref{condprop40}) binds.  As argued in the  proof of Prop. 1, the first-order condition implies that $C_\tau(\tau_2^*-\tau_1)\geq 0$ (with the c.s.), which implies that $\tau_2^*=\tau_1$ since  $C_\tau(0)=0$. Thus, when $\sigma_2^{D^*}=1$,  an increased risk of external conflict does not affect investments in fiscal capacity. 

\medskip
Now we  examine the case in which  (\ref{condprop40}) holds but  (\ref{condprop4}) is not satisfied. Note that we also have that $\gamma=0$, but this time  $\sigma_2^{D^*}=0$. Replacing these values in (\ref{tau2fin0}), and differentiating with respect to $\alpha$, we have that the sign of $\vfrac{\partial \tau^*_2}{\partial \alpha}$ depends on the sign of  $-\vfrac{\partial \phi}{\partial \alpha}-\mu$. Note from (\ref{probturnov}) that when $\gamma=0$, $\phi=(1-\alpha)\epsilon$, and $\vfrac{\partial \phi}{\partial \alpha}=-\epsilon $, with which  $-\vfrac{\partial \phi}{\partial \alpha}=\epsilon-\mu>0$ when $\epsilon(1-\sigma_2^{D^*})>\mu(1-\lambda\sigma_2^{D^*})$. Thus, in this case we have that if $\epsilon(1-\sigma_2^{D^*})>\mu(1-\lambda\sigma_2^{D^*})$, an increased risk of external conflict increases investment in fiscal capacity. 

\medskip
Finally, when  (\ref{condprop40}) is not satisfied,   $\gamma=1$ and $\sigma_2^{D^*}=0$. In this scenario, differentiating $\tau^*_2$ with respect to $\alpha$ we have from (\ref{tau2fin0}) that the sign of $\vfrac{\partial \tau^*_2}{\partial \alpha}$ depends on the sign of  $-\vfrac{\partial \phi}{\partial \alpha}$, but in this case have that $\vfrac{\partial \phi}{\partial \alpha}=\omega-\delta$, which by (\ref{phiowwp}) is strictly positive. So, $-\vfrac{\partial \phi}{\partial \alpha}=-(\omega-\delta)<0$. Thus, in this case we have that an increased risk of external conflict decreases investment in fiscal capacity. 
\end{proof} 


\hbox {} \newpage

\bibliographystyle{aer}
\bibliography{bibstatecapacity1}


\hbox {} \newpage
\appendix
\small
\subsection*{Online Appendix for ``External Threats, Political Turnover and Fiscal Capacity" by Hector Galindo-Silva}
\setcounter{equation}{0}
\renewcommand\theequation{OA.\arabic{equation}}

In this web Appendix, I extend the baseline model to include the possibility that a civil war affects the level of institutional cohesiveness. Specifically, I consider the case in which there is no redistribution of resources after a revolution, i.e., that $\sigma_2^D=0$ if $O_1$ is in power in period 2 due to a civil war. I proceed as in the main text, by first examining policy-making, then the decision to start a civil war and finally investment in fiscal capacity and political turnover. 

\subsubsection*{\emph{Policy-making}}

We distinguish between governments led by $O_1$, depending on whether $O_1$ came to power via a revolution or via an election. In both cases we still have a corner solution with $t_2=\tau_2$, but now $r_2^{I_2}$ and $r_2^{O_2}$ depend on whether $O_1$ obtained its period-2 power through a revolution. As previously mentioned, I assume that $\sigma_2^D=0$ in the revolution scenario. From (\ref{BC}) and (\ref{ro1}), it is easy to see that $r_2^{O_2}=0$ and  $r_2^{I_2}=2\tau_2w$. 

\subsubsection*{\emph{Civil war}}
First, I  re-define $O_1$'s period-2 indirect utility when $O_1$ came to power via a revolution:
\footnotesize
\begin{equation}
\label{IUIapp}
W^{O_1}(\tau_2|I_2=O_1,\sigma_2^D=0)=(1-\tau_2)w + 2\tau_2w
\end{equation}
\small
 With this change,  $O_1$'s expected utility (in period 2) in the event of a civil war is now
\footnotesize
\begin{equation}
\begin{split}
\alpha\Big[(\omega+\rho\lambda)  W^{O_1}(\tau_2|I_2=O_1,\sigma_2^D=0)+(1-\omega-\rho)W^{O_1}(\tau_2|I_2=I_1)+\rho(1-\lambda)  W^{O_1}(\tau_2|I_2=F)\Big] \\+(1-\alpha)\Big[\delta  W^{O_1}(\tau_2|I_2=O_1,\sigma_2^D=0)+(1-\delta)W^{O_1}(\tau_2|I_2=I_1)\Big]
\end{split}
\end{equation}
\small
and in the case of internal peace is
\footnotesize
\begin{equation}
\begin{split}
\alpha\Big[\mu\lambda W^{O_1}(\tau_2|I_2=O_1)+(1-\mu)W^{O_1}(\tau_2|I_2=I_1)+\mu(1-\lambda)  W^{O_1}(\tau_2|I_2=F)\Big]\\+(1-\alpha)\Big[\epsilon  W^{O_1}(\tau_2|I_2=O_1)+(1-\epsilon)W^{O_1}(\tau_2|I_2=I_1)\Big].
\end{split}
\end{equation}
\small
Rearranging, we have that $O_1$ triggers a civil war when
\footnotesize
\begin{equation}
\begin{split}
(\alpha\omega+(1-\alpha)\delta) [W^{O_1}(\tau_2|I_2=O_1,\sigma_2^D=0)- W^{O_1}(\tau_2|I_2=I_1)]+\alpha\rho\lambda W^{O_1}(\tau_2|I_2=O_1,\sigma_2^D=0)\\
-(1-\alpha)\epsilon  [W^{O_1}(\tau_2|I_2=O_1)- W^{O_1}(\tau_2|I_2=I_1)]-\alpha\mu\lambda W^{O_1}(\tau_2|I_2=O_1)\\
>\alpha[\rho -\mu ][W^{O_1}(\tau_2|I_2=I_1)- (1-\lambda)W^{O_1}(\tau_2|I_2=F)].
\end{split}
\end{equation} 
\small
Replacing (\ref{IUI}), (\ref{IUO}), (\ref{IUOF}) and (\ref{IUIapp}), and rearranging,  this condition is equivalent to
\footnotesize
\begin{equation}
\label{ }
\begin{split}
\sigma^F\big(\alpha(\rho -\mu)(1-\lambda\sigma^D)+\alpha\omega+(1-\alpha)\delta-(1-\alpha)\epsilon (1 - \sigma^D) +\alpha\rho\lambda\sigma^D\big)\\
>2\big(\alpha(\rho -\mu)(\sigma^D-\lambda) -(\alpha\omega+(1-\alpha)\delta) +(1-\alpha)\epsilon (1 - \sigma^D)-\alpha\rho\lambda\sigma^D\big)
\end{split}
\end{equation}
\small
and we still can defined a threshold for $\sigma^F$, denoted by $\overline{\sigma}^{F'}$ such that there is a civil war for all $\sigma^F>\overline{\sigma}^{F'}$, where
\footnotesize
\begin{equation}
\label{sigmaFthresholdappendx}
\overline{\sigma}^{F'}\equiv\frac{2\big(\alpha(\rho -\mu)(\sigma^D-\lambda) -(\alpha\omega+(1-\alpha)\delta) +(1-\alpha)\epsilon (1 - \sigma^D)-\alpha\rho\lambda\sigma^D\big)}{\alpha(\rho -\mu)(1-\lambda\sigma^D)+\alpha\omega+(1-\alpha)\delta-(1-\alpha)\epsilon (1 - \sigma^D) +\alpha\rho\lambda\sigma^D}
\end{equation}
\small
 provided that the denominator is positive.
 
\subsubsection*{\emph{Investment in fiscal capacity}}
As for  $I_1$'s decision to invest in fiscal capacity, $I_1$'s utility in the event of a civil war is now
\footnotesize
\begin{equation}
\label{IUOIappalt}
W^{I_1}(\tau_2|I_2=O_1,\sigma^D=0)=(1-\tau_2)m.
\end{equation}
\small
 Thus, in the event of civil war, $I_1$'s expected utility becomes
\footnotesize
\begin{equation}
\label{I1expcw}
\begin{split}
W^{I_1}(\tau_1,C(\tau_2-\tau_1))\\+\alpha\Big[(1-\omega-\rho)W^{I_1}(\tau_2|I_2=I_1)+(\omega+\rho\lambda) W^{I_1}(\tau_2|I_2=O_1,\sigma^D=0)+\rho(1-\lambda) W^{I_1}(\tau_2|I_2=F)\Big]\\
+(1-\alpha)\Big[(1-\delta)W^{I_1}(\tau_2|I_2=I_1)+\delta W^{I_1}(\tau_2|I_2=O_1,\sigma^D=0)\Big].
\end{split}
\end{equation}
\small
Differentiating (\ref{I1expcw}) with respect to $\tau_2$, and setting it equal to zero, the first-order condition is
\footnotesize
 \begin{equation}
 \begin{split}
-W_C^{I_1}(\tau_1,C(\tau^*_2-\tau_1))C_\tau(\tau^*_2-\tau_1)  \geq \\
\alpha\Big[(1-\omega-\rho)W_\tau^{I_1}(\tau^*_2|I_2=I_1)+(\omega+\rho\lambda)  W_\tau^{I_1}(\tau^*_2|I_2=O_1,\sigma^D=0)+\rho(1-\lambda) W_\tau^{I_1}(\tau^*_2|I_2=F)\Big]\\ 
 +(1-\alpha)\Big[(1-\delta)W_\tau^{I_1}(\tau^*_2|I_2=I_1)+\delta W_\tau^{I_1}(\tau^*_2|I_2=O_1,\sigma^D=0)\Big]
\end{split}
\end{equation}
\small
which, rearranging, and combining it with (\ref{IUII}), (\ref{IUOIappalt}), (\ref{IUOFI}) and (\ref{IUFP}), is equivalent to  
\footnotesize
 \begin{equation}
 \begin{split}
2C_\tau(\tau^*_2-\tau_1)\geq 2m\Big[1-(\alpha\omega+(1-\alpha)\delta+\alpha\rho)\Big]-m(1+\sigma^D)
\end{split}
\end{equation}
\small
with the complementary slackness condition associated to  $\tau^*_2-\tau_1\geq 0$.  For $\tau^*_2>\tau_1$, the complementary slackness condition implies that equality holds. Thus, solving for $\tau^*_2$, this condition can be written as 
\footnotesize
\begin{equation}
\label{tau2finapp}
 \tau^*_2 = C_\tau^{-1}\Big(m\big[-(\alpha\omega+(1-\alpha)\delta+\alpha\rho) +\vfrac{(1-\sigma^D)}{2}\big]\Big) +\tau_1.
\end{equation}
\small
Differentiating (\ref{tau2finapp}) with respect to $\alpha$ and rearranging, we have that the sign of $\frac{\partial \tau^*_2}{\partial \alpha}$ depends on the sign of
\footnotesize
\begin{equation}
\label{ }
\frac{\partial }{\partial \alpha}\Big(-(\alpha\omega+(1-\alpha)\delta+\alpha\rho)+\vfrac{(1-\sigma^D)}{2}\Big)=-(\omega-\delta+\rho)
\end{equation}
which by (\ref{phiowwp}) is strictly negative. 
\small
The case of no civil war is similar to that developed in the main text, the outcome of which is summarized in Prop. 2.B. Thus, we can formulate alternative versions of Props. 1 and 2, which establish conditions for a change in political turnover and in investment in fiscal capacity as a consequence of an increased risk of external conflict. 

  \begin{prop1a} Consider the above-described game. Then, there is a unique threshold for $\sigma^F$, denoted by $\overline{\sigma}^{F'}$ and given by (\ref{sigmaFthresholdappendx}), such that in equilibrium: 
  \begin{itemize} \itemsep-0.1em 
      \item[(1a.A)]  If $\sigma^F>\overline{\sigma}^{F'}$, an increased risk of external conflict increases  $\phi$. 
       \item[(1a.B)] If $\sigma^F\leq\overline{\sigma}^{F'}$, an increased risk of external conflict decreases  $\phi$. 
\end{itemize}
 \end{prop1a}

  \begin{prop2a} Consider the above-described game. Then, there is a unique threshold for $\sigma^F$, denoted by $\overline{\sigma}^{F'}$ and given by (\ref{sigmaFthresholdappendx}), such that in equilibrium: 
  \begin{itemize} \itemsep-0.1em 
      \item[(2a.A)]  If $\sigma^F>\overline{\sigma}^{F'}$, then an increase in the risk of an external conflict decreases investments in fiscal capacity. 
       \item[(2a.B)] If $\sigma^F\leq\overline{\sigma}^{F'}$, then: 
       		\begin{itemize}\itemsep-0.1em 
  			\item[(2a.B.1)] If $(\epsilon-\mu)>\sigma^D(\epsilon-\lambda\mu)$,  an increase in the risk of an external conflict increases investments in fiscal capacity.
			\item[(2a.B.2)] If $(\epsilon-\mu)=\sigma^D(\epsilon-\lambda\mu)$,  an increase in the risk of an external conflict does not affect investments in fiscal capacity. 
 		 	\item[(2a.B.3)] If $(\epsilon-\mu)< \sigma^D(\epsilon-\lambda\mu)$,  an increase in the risk of an external conflict decreases investments in fiscal capacity. 
		\end{itemize}
\end{itemize}
 \end{prop2a}
 
Comparing (\ref{sigmaFthreshold}) and (\ref{sigmaFthresholdappendx}), note that  $\overline{\sigma}^{F}>\overline{\sigma}^{F'}$. The intuition for this result is straightforward: in this extension of the baseline model, $O_1$ has a greater incentive to start a civil war, so the conditions for more investment in fiscal capacity are more difficult to satisfy. 

\bigskip

Finally, I state the alternative version of Prop 3. The proof is the  same as that for Prop 3. 

   \begin{prop3a} Consider the above-described game. Then, there is a unique threshold for $\sigma^F$, denoted by $\overline{\sigma}^{F'}$ and given by (\ref{sigmaFthresholdappendx}), such that in equilibrium: 
    \begin{itemize} \itemsep0em 
  \item[(3a.A)] If $\sigma^F>\overline{\sigma}^{F'}$, an increased risk of external conflict  implies an increase in $\phi$ and decreased investment in fiscal capacity. 
  \item[(3a.B)] If $\sigma^F\leq\overline{\sigma}^{F'}$, then: 
       		\begin{itemize}\itemsep-0.1em 
  			\item[(3a.B.1)] If $(\epsilon-\mu)>\sigma^D(\epsilon-\lambda\mu)$,  an increased risk of external conflict implies a decrease in $\phi$ and increased investment in fiscal capacity. 
			\item[(3a.B.2)] If $(\epsilon-\mu)\leq\sigma^D(\epsilon-\lambda\mu)$,  an increased risk of external conflict implies a decrease in $\phi$ but not increased investment in fiscal capacity. 
		\end{itemize}  
  \end{itemize}
    \end{prop3a}


\end{document}